\documentclass[12pt]{article}
\usepackage{epsf,amssymb,psfrag}

\catcode`\@=11
\def \thesection {\arabic{section}.}

\def \be  {\begin{equation}}
\def \ee  {\end{equation}}
\def \ba  {\begin{eqnarray}}
\def \ea  {\end{eqnarray}}
\def \baa {\begin{eqnarray*}}
\def \eaa {\end{eqnarray*}}
\def \bb  {\begin {thebibliography} }
\def \eb  {\end{thebibliography}}
\def \lab #1 {\label{#1}}

\newcommand\re[1]{({\ref{#1}})}

\def \qqqquad {\qquad\qquad}
\def \matrix #1 {\left(\begin{array}{cc} #1 \end{array}\right)}

\def \tr {\mathop{\rm tr}\nolimits}
\def \Im {\mathop{\rm Im}\nolimits}
\def \Re {\mathop{\rm Re}\nolimits}

\def \e  {\mathop{\rm e}\nolimits}
\newcommand\lr[1]{{\left({#1}\right)}}

\newcommand \vev [1] {\langle{#1}\rangle}

\newcommand \ket [1] {|{#1}\rangle}

\newcommand{\as}{\ifmmode\alpha_{\rm s}\else{$\alpha_{\rm s}$}\fi}
\newcommand{\asbar}{\ifmmode\bar{\alpha}_{\rm s}\else{$\bar{\alpha}_{\rm s}$}\fi}

\font\cmss=cmss12 
\def\inbar{\,\vrule height1.5ex width.4pt depth0pt}
\def\IC{\relax\hbox{$\inbar\kern-.3em{\rm C}$}}
\def\IZ{\relax{\hbox{\cmss Z\kern-.4em Z}}}
\def\IR{{\hbox{{\rm I}\kern-.2em\hbox{\rm R}}}}

\def\IP{{\hbox{{\rm I}\kern-.2em\hbox{\rm P}}}}
\def\II{\hbox{{1}\kern-.25em\hbox{l}}}

\def\numberbysection{\@addtoreset{equation}{section}
                     \def\theequation{\thesection\arabic{equation}}}
\numberbysection

\newcommand \mybf[1] {\mbox{\boldmath$\scriptstyle {#1} $}}
\newcommand \Mybf[1] {\mbox{\boldmath$ {#1} $}}

\psfrag{a1}[cl][cc]{$s+iy$}
\psfrag{a2}[cl][cc]{$s-iy$}
\psfrag{a3}[cl][cc]{$s-ix$}
\psfrag{a4}[cl][cc]{$s+ix$}
\psfrag{a5}[bc][cc][1][90]{$i(y-x)$}
\psfrag{b1}[cr][cc]{$s+iy$}
\psfrag{b2}[cr][cc]{$s-iy$}
\psfrag{b3}[cr][cc]{$s-ix$}
\psfrag{b4}[cr][cc]{$s+ix$}

\begin{document}

\begin{titlepage}

\begin{flushright}
\begin{tabular}{l}
LPT--Orsay--02--88\\
RUB-TP2-13/02\\
hep-th/0210216
\end{tabular}
\end{flushright}


\vskip3cm

\begin{center}
  {\large \bf Separation of  variables for the quantum $SL(2,{\mathbb R })$
    spin chain }

\def\thefootnote{\fnsymbol{footnote}}%
\vspace{1cm}
{\sc S.\'{E}. Derkachov}${}^1$, {\sc G.P.~Korchemsky}${}^2$
 and {\sc A.N.~Manashov}${}^3$\footnote{ Permanent
address:\ Department of Theoretical Physics,  Sankt-Petersburg State University,
St.-Petersburg, Russia}
\\[0.5cm]

\vspace*{0.1cm} ${}^1$ {\it
Department of Mathematics, St.-Petersburg Technology Institute,\\
St.-Petersburg, Russia
                       } \\[0.2cm]
\vspace*{0.1cm} ${}^2$ {\it
Laboratoire de Physique Th\'eorique%
\footnote{Unite Mixte de Recherche du CNRS (UMR 8627)},
Universit\'e de Paris XI, \\
91405 Orsay C\'edex, France
                       } \\[0.2cm]
\vspace*{0.1cm} ${}^3$
 {\it
Institut f\"ur Theoretische Physik II, Ruhr-Universit\"at Bochum,\\
D-44780 Bochum, Germany}

\vskip2cm
{\bf Abstract:\\[10pt]} \parbox[t]{\textwidth}{
We construct a representation of the Separated Variables (SoV) for the quantum
$SL(2,\mathbb{R})$ Heisenberg closed spin chain following the Sklyanin's approach
and obtain the integral representation for the eigenfunctions of the model. We
calculate explicitly the Sklyanin measure defining the scalar product in the SoV
representation and demonstrate that the language of Feynman diagrams is extremely
useful in establishing various properties of the model. The kernel of the unitary
transformation to the SoV representation is described by the same ``pyramid
diagram'' as appeared before in the SoV representation for the $SL(2,\mathbb{C})$
spin magnet. We argue that this kernel is given by the product of the Baxter
$\mathbb{Q}-$operators projected onto a special reference state.}
\vskip1cm

\end{center}

\end{titlepage}


\newpage
\setcounter{footnote}{0}

\section{Introduction}
\label{intr}

In this paper we study the spectral problem for quantum $SL(2,\mathbb{R})$ spin
magnet within the Quantum Inverse Scattering Method~\cite{QISM,XXX,ABA}. The
model is a generalization of the famous spin$-1/2$ Heisenberg XXX chain. It
describes the nearest neighbour interaction between $N$ spins with the
corresponding spin operators being the generators of infinite-dimensional
representation of the $SL(2,\mathbb{R})$ group.

The interest to studying noncompact spin magnets is twofold. From the one side,
there exists a deep relation between integrable models and quantum
$(3+1)-$dimensional Yang-Mills theories~\cite{GKK}. It turns out that the scale
dependence of certain correlation functions on the light-cone, like those
defining a baryon distribution amplitude and twist-three light-cone
distributions, is governed by the evolution equations which upon redefinition of
the variables coincide with the Schr\"odinger equation for the $SL(2,\mathbb{R})$
spin chain~\cite{BDM,BDKM,AB}. The number of sites in the lattice model is equal
to the number of fields entering the correlation functions. The spin operators
are the generators of the so-called collinear $SL(2,\mathbb{R})$ subgroup of the
full conformal group of the classical Yang-Mills Lagrangian. Another example
comes from the studies of high-energy (Regge) asymptotics of the scattering
amplitudes in multi-colour QCD. As was shown in~\cite{L1,FK}, the spectrum of
multi-gluonic compound states responsible for a power rise with the energy of the
scattering amplitudes is described by the $SL(2,{\mathbb C})$ spin magnet.

{}From the other side, an exact solution of the spectral problem for completely
integrable quantum mechanical systems with infinitely dimensional quantum space,
like periodic Toda chain and noncompact $SL(2)$ spin magnet, represents a
challenge for the theory of Integrable Models. The conventional methods like the
Algebraic Bethe Ansatz~\cite{QISM,ABA} (ABA) are not always applicable to such
systems and one has to rely instead on the methods of the Baxter $\mathbb
Q-$operator~\cite{Baxter} and the Separation of Variables (SoV)~\cite{Sklyanin}.
Being combined together, the two methods allow us to determine the energy
spectrum of the model in terms of the eigenvalues of the $\mathbb Q-$operator and
construct an integral representation for the corresponding eigenfunctions by
going over to the representation of the separated variables.

In spite of the fact that the methods have been formulated a long time ago, the
explicit construction of the $\mathbb Q-$operator and the unitary transformation
to the SoV representation for each particular model remains an extremely
nontrivial task (for some known examples see review \cite{Skl1} and references
therein). It is equally nontrivial to solve the emerging functional relations
(the Baxter equations) and reconstruct a fine structure of the spectrum.

A powerful algebraic approach to constructing the SoV representation has been
developed by Sklyanin~\cite{Sklyanin}. It allows one to establish the relations
both for the eigenfunctions in the separated coordinates and for the scalar
product in the SoV representation. Their solutions are defined up to
multiplication by an arbitrary periodic function. For models with
finite-dimensional quantum space, like conventional $SU(2)$ Heisenberg magnet,
the separated coordinates take a discrete, finite set of values and, as a
consequence, the above relations can be uniquely solved. For models with
infinite-dimensional quantum space, like $SL(2,\mathbb{R})$ Heisenberg magnet and
periodic Toda chain, the separated coordinates take continuous values and the
question arises how to fix the ambiguity or, equivalently, what are the
additional conditions that one has to impose on the eigenfunctions and the
integration measure in the SoV representation. To answer this question within the
Sklyanin's approach, one has to provide an explicit construction of the unitary
transformation to the SoV representation and identify the analytical properties
of the resulting expressions for the eigenfunctions in the separated coordinates.

For the periodic Toda chain this program has been carried out by Kharchev and
Lebedev in the series of papers~\cite{KL}. The approach proposed in
Ref.~\cite{KL} is based on the relation between the wave functions of the
periodic and open Toda chains originating from a specific form of the Lax
operator. Notice that similar relation does not hold for the $SL(2)$ magnet.
In the present paper we construct the SoV representation for
the $SL(2,\mathbb{R})$ spin chain following the method developed in~\cite{DKM,VL}
in application to the  $SL(2,{\mathbb C})$ spin magnet. The method is general
enough as it is applicable to both the $SL(2,\mathbb{R})$ spin chain and the Toda
chain.
%
%
%

Our main results include: calculation of the Sklyanin's measure defining the
scalar product in the SoV representation, establishing the relation between the
kernel of the unitary transformation to the SoV representation and the Baxter
$\mathbb{Q}-$operator constructed in~\cite{SD}, proof of the equivalence of the
ABA and the SoV method for the $SL(2,\mathbb{R})$ spin chain.

The paper is organized as follows. In Section~2 we define the model and describe
its integrability properties. In Section~3 we apply the Sklyanin approach and
present an explicit construction of the unitary transformation to the Separated
Variables for the $SL(2,\mathbb{R})$ closed spin chain. We demonstrate that the
kernel of the SoV transformation admits a simple interpretation in terms of
Feynman diagrams. In Section~4 we follow the diagrammatical approach and derive
the invariant scalar product in the SoV representation. In Section~5 we establish
the relation between the transition function to the SoV representation and the
Baxter $\mathbb{Q}-$operator. It allows us to obtain the expressions for the
eigenfunctions of the model in the separated variables. Section~5 contains
concluding remarks. Some details of the calculations can be found in Appendix~A.
In Appendix~B we argue that the Algebraic Bethe Ansatz and the SoV method lead to
the same expressions for the eigenstates of the $SL(2,\mathbb{R})$ spin magnet.

\section{ The quantum $SL(2,\mathbb{R})$ spin magnet }
\label{II}

The quantum $SL(2,\mathbb{R})$ spin magnet is a one-dimensional lattice model of
$N$ interacting spins $\vec S_n=(S_n^{0}\,,S_n^{+}\,,S_n^{-})$ (with
$n=1,...,N$). The spin operators in different sites commute with each other and
obey the standard $sl(2)$ commutation relations
\begin{equation}
\label{comm-r}
[S_n^0,S_n^\pm]=\pm\,\hbar\, S_n^\pm\,, \qquad [S_n^+,S_n^-]=2\hbar\, S_n^0\,.
\end{equation}
The corresponding quadratic Casimir operator is defined as
\be
\vec S_n^2=S_n^0S_n^0+\frac12(S_n^+ S_n^-+S_n^- S_n^+)=\hbar^2\,s_n(s_n-1)\,.
\ee
We shall impose periodic boundary conditions, $\vec S_{N+1}\equiv\vec S_1$, and
put $\hbar=1$ for simplicity. In addition, we shall assume that the spin chain is
homogenous, $s_1=...=s_N=s$. Generalization to the case of inhomogeneous spin
chain will be discussed in Section~6.

\subsection{Hamiltonian of the model}

The Hamiltonian of the homogenous $SL(2,\mathbb{R})$ spin magnet is defined
as~\cite{XXX,BT}
\begin{equation}
\label{H}
\mathcal{H}_N=\sum_{n=1}^N H_{n,n+1}\,,\qquad H_{n,n+1}=
\psi(J_{n,n+1})-\psi(2s)
\,,
\end{equation}
where $\psi(z)=d \ln\Gamma(z)/dz$ is the Euler $\psi$-function. The operator
$J_{n,n+1}$ is related to the sum of two neighbouring spins
\begin{equation}
\label{J}
J_{n,n+1}(J_{n,n+1}-1)=(\vec{S}_n+\vec{S}_{n+1})^2\,,
\end{equation}
$J_{N,N+1}=J_{N,1}$ and its eigenvalues satisfy the condition $J_{n,n+1}\ge 1/2$.
The Hamiltonian \re{H} has been constructed in Ref.~\cite{XXX} as a
generalization of the spin$-1/2$ XXX Heisenberg spin chain to high-dimensional
representations of the $SU(2)$ group. By the construction, $\mathcal{H}_N$
possesses a set of mutually commuting integrals of motion that we shall denote as
$\Mybf{q}=(q_2,...,q_N)$. Their number is large enough for the model to be
completely integrable. The definition of the operators $\Mybf{q}$ will be given
below (see Eq.~\re{tu-exp}).

In what follows, we shall use a particular representation for the spin operators
\begin{equation}
\label{s-rep}
S_n^{+}~=~iz_n^2p_n+2 s\,z_n,\ \ \ S_n^{-}~=~-ip_n,\ \ \ S_n^{0}~=~iz_np_n+ s\,,
\end{equation}
where $p_n=-i\,\partial/\partial z_n$ and $[x_n,p_m]=i\,\delta_{nm}$. The spin
$s$ is assumed to be real and $s\ge 1/2$. In this representation, the
$SL(2,\mathbb{R})$ spin magnet \re{H} can be interpreted as a one-dimensional
quantum mechanical model of $N$ interacting particles with the coordinates $z_n$
and the conjugated momenta $p_n$ ($n=1,...,N$). At $N=3$ this model has appeared
in high-energy QCD as describing the spectrum of the anomalous dimensions of the
baryon distribution amplitudes~\cite{BDKM}.

The spin operators \re{s-rep} act on the Hilbert space, $V_n$, of functions
$\Psi(z_n)\in V_n$ holomorphic in the upper half-plane
\footnote{One can choose instead $\Psi(z_n)$ to be holomorphic in the
lower half-plane. As we will show below, the two cases, $\Im z>0$ and $\Im z<0$,
correspond to the different values of the total momentum of the system, $p>0$ and
$p<0$, respectively.}, $\Im z_n >0$, and normalizable with respect to the scalar
product~\cite{Gelfand}
\begin{equation}
\label{norm}
\|\Psi\|^2~=~\int_{\Im z>0}\mathcal{D}z\, |\Psi(z)|^2\,,
\end{equation}
with $z=x+iy$. Here integration is performed over the upper half-plane  and the
integration measure isdefined as%
\footnote{Performing the conformal mapping $w=i(z-i)/(z+i)$, one can bring
this expression to  a canonical form involving the integration over an interior
of the unit disk in the $w-$plane~\cite{Gelfand}.}
\be
\mathcal{D}z=\frac{2s-1}{\pi}\,d^2z \,
(2\Im z)^{2s-2}=\frac{2s-1}{\pi}dxdy\,(2y)^{2s-2}\,.
\label{measure}
\ee
The spin $s$ in Eqs.~\re{s-rep} and \re{measure} takes arbitrary real values
$s\ge 1/2$. For $s$ integer or half integer, the Hilbert space $V_n$ coincides
with the linear space of the unitary irreducible representation of the
$SL(2,\mathbb{R})$ group of the discrete series~\cite{Gelfand}
$$
\left [T(g^{-1})\Psi\right ](z)~=~\frac{1}{(cz+d)^{2s}}\,\Psi\left(\frac{az+b}{cz+d}\right)\,,
\qquad g=\lr{\begin{array}{cc}
  a & b \\
  c & d \\
\end{array}}\in SL(2,\mathbb{R})\,.
$$

In this paper we shall study the Schr\"odinger equation for the Hamiltonian
\re{H}
\be
\mathcal{H}_N
\Psi_{\mybf{q}}(z_1,...,z_N)=E_{\mybf{q}}\,\Psi_{\mybf{q}}(z_1,...,z_N)\,.
\label{Sch}
\ee
The Hamiltonian $\mathcal{H}_N$ acts on the quantum space of the model
$\mathcal{V}_N$, given by the direct product of the Hilbert spaces in each site,
$\mathcal{V}_N=\prod_{n=1}^N\otimes V_n$. In distinction with conventional
(compact) Heisenberg spin magnet, $\mathcal{V}_N$ is infinite-dimensional for
arbitrary finite $N$. The eigenstates
$\Psi_{\mybf{q}}(z_1,...,z_N)\in\mathcal{V}_N$ are holomorphic functions of the
$z-$coordinates in the upper half-plane, $\Im z_n
>0$ for $n=1,...,N$, normalizable with respect to the scalar product
\be
\|\Psi_N\|^2~=~\int\mathcal{D}^Nz\, |\Psi(z_1,...,z_N)|^2\,.
\label{scalar}
\ee
Here $\int\mathcal{D}^Nz=\prod_{n=1}^N \int_{\Im z_n>0}\mathcal{D}z_n$ and the
measure $\mathcal{D}z_n$ is given by \re{measure}. In Eq.~\re{Sch} we indicated
explicitly the dependence on the integrals of motion $\Mybf{q}=(q_2,...,q_N)$.
The Hamiltonian \re{H} commutes with the total spin of the magnet $\vec S=\vec
S_1+...+\vec S_N$ and one of its component, $iS_-=p_1+...+p_N$, defines the total
momentum of the system. This allows us to assign a definite value of the momentum
to the solutions to \re{Sch}, $iS_-\Psi_{\mybf{q},p}=p\,\Psi_{\mybf{q},p}$,
leading to
\be
\Psi_{\mybf{q},p}(z_1,...,z_N) = \int_{-\infty}^\infty d x_0 \, \e^{ip\, x_0}
\Psi_{\mybf{q}}(z_1-x_0,...,z_N-x_0)\,,
\ee
where integration goes along the real $x-$axis. Notice that in virtue of the
$SL(2)$ invariance of the Hamiltonian, $[\mathcal{H}_N,\vec S]=0$, the energy
$E_{\mybf{q}}$ does not depend on $p$.

It is straightforward to verify that for arbitrary real $s$ the spin operators
\re{s-rep} are anti-hermitian with respect to the scalar product \re{norm} and
\re{scalar}
\be
\lr{S_n^0}^{\dagger} = - S_n^0\,,\qquad
\lr{S_n^\pm}^{\dagger} = - S_n^\pm\,
\label{anti-h}
\ee
This property ensures that the Hamiltonian \re{H} and the total momentum operator
$iS_-$ are hermitian on the Hilbert space of the model. As a consequence, the
energy $E_{\mybf{q}}$ and the total momentum $p$ take real values. As we will
show in Section~3, the property \re{anti-h} plays an important r\^ole in our
construction of the SoV representation.

\subsection{Integrals of motion}

To construct the integrals of motion of the model, one follows the $R-$matrix
approach~\cite{QISM,ABA,XXX}. The Lax operator for the $SL(2,\mathbb{R})$ magnet
is defined as
\begin{equation}
\label{Lax}
L_n(u)=u+i (\vec\sigma\cdot  \vec{S}_n)=
\left(\begin{array}{cc}u+iS^0_n& iS^-_n\\iS^+_n&u-iS^0_n\end{array}\right)\,.
\end{equation}
It acts on the direct product of the auxiliary space and the quantum space in the
$n$th site, ${\mathbb C}^2\otimes V_n$, and satisfies the Yang-Baxter commutation
relations involving a rational $R-$matrix~\cite{QISM,ABA,XXX}. Taking the product
of $N$ Lax operators in the auxiliary space, one obtains the monodromy matrix
\begin{equation}
\label{Tu}
{\mathbb T}_N(u)=L_1(u)\ldots L_N(u)=
\left(\begin{array}{cc}A_N(u)&B_N(u)\\C_N(u)&D_N(u)\end{array}\right),
\end{equation}
with the operators $A_N(u)$, $...$, $D_N(u)$ acting on the quantum space of the
model $\mathcal{V}_N$. They satisfy the following Yang-Baxter
relations~\cite{QISM}
\begin{eqnarray}
B_N(u)\,B_N(v)&=&B_N(v)\,B_N(u)\,,\nonumber\\
(v-u+i)\, A_N(v)\,B_N(u)&=&(v-u)\,B_N(u)\,A_N(v)~+~i\,A_N(u)\,B_N(v)\,,\label{AB}\\
(v-u-i)\, D_N(v)\,B_N(u)&=&(v-u)\,B_N(u)\,D_N(v)~-~i\,D_N(u)\,B_N(v)\nonumber\,.
\end{eqnarray}
The quantum determinant of the monodromy matrix \re{Tu} is given by~\cite{IK}
\begin{equation}
\det{}_{\rm q} \mathbb{T}_N(u)=A_N(u)\,D_N(u+i)-C_N(u)\,B_N(u+i)=(u+is)^N\,(u+i-is)^N\,.
\label{q-det}
\end{equation}
It follows from the Yang-Baxter relations that the auxiliary transfer matrix
$\widehat t_N(u)$, defined as a trace of the monodromy matrix over the auxiliary
space
\begin{equation}
\label{tu}
\widehat t_N(u)=\tr\,{\mathbb T}_N(u)=A_N(u)+D_N(u)\,,
\end{equation}
commutes with itself for different values of the spectral parameter, with the
Hamiltonian of model \re{H} and the operator of the total spin
\begin{equation}
\label{c-tsh}
[\widehat t_N(u),\widehat t_N(v)]=[\widehat t_N(u),\mathcal{H}_N]=[\widehat
t_N(u),\vec{S}]=0\,.
\end{equation}
Substituting \re{Tu} into \re{tu} and taking into account the explicit form of
the Lax operator \re{Lax} one finds that $\widehat t_N(u)$ is a polynomial of
degree $N$ in the spectral parameter $u$ with the operator valued coefficients
\begin{equation}
\label{tu-exp}
\widehat t_N(u)=2u^N+\widehat q_2\,u^{N-2}+\ldots+\widehat q_N\,.
\end{equation}
One deduces from \re{c-tsh} that the operators $\widehat q_2$, $...$, $\widehat
q_N$ form a family of mutually commuting, $SL(2)$ invariant integrals of motion.
Together with the total momentum of the system $p$, their eigenvalues
$\Mybf{q}=(q_2,...,q_N)$ form a complete set of the quantum numbers specifying
the solutions to the Schr\"odinger equation \re{Sch}. As a consequence, the
spectral problem \re{Sch} can be reformulated as
\begin{equation}
\label{ei-tu}
\widehat t_N(u)\,\Psi_{\mybf{q},p}(z_1,\ldots,z_N)~=~
t_N(u)\,\Psi_{\mybf{q},p}(z_1,\ldots,z_N)\,,
\end{equation}
with $t_N(u)$ being an eigenvalue of the transfer matrix \re{tu-exp}. The
Hamiltonian of the model, Eq.~\re{H}, can be obtained in a similar manner from
the fundamental transfer matrix, which is constructed analogously to \re{tu} and
\re{Tu} from the Lax operators acting on the direct product of two copies of the
quantum space $V\otimes V$~\cite{XXX}.

As follows from their definition, Eqs.~\re{tu-exp}, \re{tu} and \re{Lax}, the
integrals of motion $\widehat q_k$ take a form of polynomials in the spin
operators $\vec S_n$ (with $n=1,...,N$) of degree $k$. For instance,
\begin{equation}\label{q2}
\widehat q_2 = -2\sum_{m>n}(\vec{S}_m\cdot \vec{S}_n) = -\vec S^2 + Ns(s-1)\,,
\end{equation}
with $\vec S^2$ being the Casimir operator corresponding to the total spin of $N$
particles. In the representation \re{s-rep}, $\widehat q_k$ is given by a
complicated $k$th order differential operator and Eq.~\re{ei-tu} leads to the
system of $(N-1)-$differential equations $(\widehat q_k
-q_k)\Psi_{\mybf{q},p}\,(z_1,\ldots,z_N)=0$ with $k=2,\ldots,N$. Its exact
solution for arbitrary $N$ becomes problematic. To overcome this difficulty we
shall apply the method of Separated Variables (SoV) developed by Sklyanin
in~\cite{Sklyanin}.

\section{Separation of Variables for the $SL(2,\mathbb{R})$ magnet}

Let us  construct an integral representation for the eigenfunctions of the
quantum $SL(2,\mathbb{R})$ magnet, $\Psi_{\mybf{q},p}\,(z_1,\ldots,z_N)$, by
going over from the coordinate $z-$representation to the representation of the
Separated
Variables $(p,\Mybf{x})=(p,x_1,...,x_{N-1})$ 
\be
\Psi_{\mybf{q},p}\,(z_1,\ldots,z_N)=\int d^{N-1}\Mybf{x}\,\mu(\Mybf{x})\,
U_{p,\mybf{x}}(z_1,\ldots,z_N)\,
\Phi_{\mybf{q}}(\Mybf{x})\,,
\label{SoV-gen}
\ee
where $\mu(\Mybf{x})$ is the integration measure on the $\Mybf{x}-$space and
integration region will be specified below. $U_{p,\mybf{x}}(z_1,\ldots,z_N)$ is
the kernel of the unitary operator corresponding to this transformation
\be
U_{p,\mybf{x}}(z_1,\ldots,z_N)=\vev{z_1,\ldots,z_N|p,\Mybf{x}}\,,
\label{U-ker}
\ee
where we introduced standard notations for the bra- and ket-vectors on the
quantum space of the model. $\Phi_{\mybf{q}}(\Mybf{x})$ is the eigenfunction in
the separated coordinates
\be
\Phi_{\mybf{q}}(\Mybf{x})\,\delta(p-p') = \vev{p',\Mybf{x}|\Psi_{\mybf{q},p}}
=\int {\cal D}^N z\, \lr{U_{p',\mybf{x}}
(z_1,\ldots,z_N)}^*\Psi_{\mybf{q},p}\,(z_1,\ldots,z_N)\,.
\label{SoV-inv}
\ee
A unique feature of the SoV representation is that the eigenfunction
$\Phi_{\mybf{q}}(\Mybf{x})$ is factorized into a product of functions depending
on a single variable 
$\Phi_{\mybf{q}}(\Mybf{x})\sim Q(x_1)\ldots Q(x_{N-1})$.

In this Section, we shall obtain an explicit expression for the transition
function to the SoV transformation, $U_{p,\mybf{x}}(z_1,\ldots,z_N)$. The
integration measure $\mu(\Mybf{x})$ and the properties of the eigenfunctions
$\Phi_{\mybf{q}}(\Mybf{x})$ will be discussed in Sections~4 and 5, respectively.

\subsection{Basics of the SoV method}
\label{Bas}

To construct the unitary transformation to the SoV representation,
Eq.~\re{U-ker}, we apply the Sklyanin's approach~\cite{Sklyanin}. In this
approach, the basis vectors $\ket{p,\Mybf{x}}$ entering \re{U-ker} are defined as
eigenstates of the operator $B_N(u)$. We recall that $B_N(u)$ was introduced in
\re{Tu} as an off-diagonal component of the monodromy operator. From the
Yang-Baxter relations for the monodromy operator follows that
$[B_N(u),B_N(v)]=[B_N(u),S_-]=0$. This allows one to define the eigenfunctions of
the operator $B_N(u)$ in such a way that they are $u-$independent and diagonalize
the operator of the total momentum $(iS_--p)\ket{p,\Mybf{x}}=0$.

One finds from \re{Tu} and \re{Lax} that $B_N(u)$ is a polynomial in $u$ of
degree $N-1$ with operator valued coefficients, $B_N(u)=iS_-\,u^{N-1}+\ldots$. It
eigenvalues can be specified by the total momentum $p$ and by the set of zeros
$\Mybf{x}=(x_1,\ldots,x_{N-1})$
\be
B_N(u)\,U_{p,\mybf{x}}(z_1,\ldots,z_N)=p\,(u-x_1)...(u-x_{N-1})\,
U_{p,\mybf{x}}(z_1,\ldots,z_N)\,.
\label{B-pol}
\ee
Notice that this relation is equivalent to the following system of equations
\begin{equation}
iS_-\,U_{p,\mybf{x}}(z) = p\,U_{p,\mybf{x}}(z)\,,\qquad
B_N(x_k)\,U_{p,\mybf{x}}(z)=0,\qquad (k=1,...N-1)\,,
\label{BxN}
\end{equation}
where $z=(z_1,\ldots,z_N)$. The solutions to Eqs.~\re{B-pol} and \re{BxN} are
defined up to an overall normalization factor. It is convenient to choose it in
such a way that $U_{p,\mybf{x}}(z_1,\ldots,z_{N-1})$ will be symmetric under
permutations of any pair of the separated coordinates
\be
U_{p,\cdots x_n \cdots x_m \cdots}(z_1,\ldots,z_{N-1})=U_{p,\cdots x_m \cdots x_n
\cdots}(z_1,\ldots,z_{N-1})\,.
\label{Weyl}
\ee
Using \re{Lax}, \re{Tu} and \re{anti-h}, one can verify that $B_N(u)$ is a
self-adjoint operator on the Hilbert space of the model for real $u$
\be
\lr{B_N(u)}^\dagger = B_N(u^*)\,.
\label{B-self}
\ee
This implies that the parameters $\Mybf{x}$ parameterizing its eigenvalues,
Eq.~\re{B-pol}, take \textit{continuous real} values. The set of corresponding
eigenstates $\ket{p,\Mybf{x}}$ is complete on the quantum space of the model
$\mathcal{V}_N$ and their orthogonality condition looks as
\ba
\vev{p',\Mybf{x'}|p,\Mybf{x}} &=& \int \mathcal{D}^N z\,
U_{p,\mybf{x}}(z_1,...,z_N) \lr{U_{p',\mybf{x'}}(z_1,...,z_N)}^*
\nonumber
\\
&=& 
\delta(p-p') \left\{\delta(\Mybf{x}-\Mybf{x'})+ \cdots \right\}
\frac{\mu^{-1}(\Mybf{x})}{(N-1)!}\,,
\label{U-ort}
\ea
where $\delta(\Mybf{x}-\Mybf{x'})\equiv
\prod_{k=1}^{N-1}\delta(x_k-x'_k)$ and ellipses denote the sum of terms with
all possible permutations inside the set $\Mybf{x}=(x_1,\ldots,x_{N-1})$.

The property \re{B-self} also holds for the operators $A_N(u)$, $D_N(u)$,
$C_N(u)$ and 
for the transfer matrix, $(\widehat t_N(u))^\dagger=\widehat t_N(u^*)$. Applying
the both sides of the Yang-Baxter relations \re{AB} to
$U_{p,\mybf{x}}(z_1,...,z_N)$ and taking $v=x_k$, one finds
\be
A_N(x_k)\,U_{p,\,\mybf{x}}(z)=a_k(\Mybf{x})\, U_{p,\,\mybf{x}+i
\mybf{e}_k}(z)\,,\qquad D_N(x_k)\,U_{p,\mybf{x}}(z)=d_k(\Mybf{x})\,
U_{p,\,{ {\mybf{x}}}-i {\mybf{e}}_k}(z_i)
\,.
\label{shiftD}
\ee
Here the notation was introduced for the unit vectors $\Mybf{e_k}$ in the
$(N-1)-$dimensional $\Mybf{x}-$space, such that $(\Mybf{e_k})_n=\delta_{nk}$ and
$\Mybf{x}+i\Mybf{e}_k\equiv (x_1,\ldots,x_k+i,\ldots,x_{N-1})$. Substituting
$u=x_k$ into \re{q-det} and applying \re{shiftD}, one finds that
$a_k(\Mybf{x})\,d_k(\Mybf{x}+i\Mybf{e}_k) =(x_k+is)^N\,(x_k+i-is)^N$. The
coefficients $a_k(\Mybf{x})$ and $d_k(\Mybf{x})$ depend on the normalization of
$U_{p,\,\mybf{x}}(z)$, or equivalently on the definition of the integration
measure in \re{U-ort}. It is convenient to normalize $U_{p,\mybf{x}}(z)$ in such
a way that
\be
a_k(\Mybf{x})=\Delta_{+}(x_k)=(x_k+is)^N\,,\qquad
d_k(\Mybf{x})=\Delta_{-}(x_k)=(x_k-is)^N\,.
\label{coef-ad}
\ee
In Eq.~\re{shiftD} we have tacitly assumed that the function
$U_{p,\,\mybf{x}}(z)$ can be analytically continued from real $\Mybf{x}$ into a
finite strip in the complex plane. We will verify this property {\it a
posteriori\/}.

Using the solutions to \re{B-pol} and \re{BxN}, one can decompose an arbitrary
state on $\mathcal{V}_N$ over the basis of the functions
$U_{p,\mybf{x}}(z_1,\ldots,z_N)$. For the eigenstates of the model, the
decomposition takes the form \re{SoV-gen} and \re{SoV-inv}. To obtain the wave
function in the separated coordinates, $\Phi_{\mybf{q}}(\Mybf{x})$, one has to
examine the action of the transfer matrix on the function
$U_{p,\mybf{x}}(z_1,\ldots,z_N)$. According to \re{tu-exp}, $\widehat t_N(u)$ is
a polynomial of degree $N$ in $u$ with the coefficient in front of $u^{N-1}$
equal to zero. As a result, it can be reconstructed from its values at $N-1$
distinct points $\widehat t_N(x_k)=A_N(x_k)+D_N(x_k)$, with $k=1,\ldots,N-1$,
using the Lagrange interpolation formula. Applying \re{shiftD} one gets
\ba
\widehat t_N(u)\,U_{p,\,\mybf{x}}(z)
&=&2(u+\sum_{k=1}^{N-1} x_k)\prod_{j=1}^{N-1}
(u-x_j)U_{p,\,\mybf{x}}(z)
\nonumber
\\
&+&
\sum_{k=1}^{N-1}
\prod_{j\neq k}\frac{u-x_j}{x_k-x_j}\bigg[\Delta_+(x_k)U_{p,\,\mybf{x}+i\mybf{e_k}}(z)
+\Delta_-(x_k)U_{p,\,\mybf{x}-i\mybf{e_k}}(z)\bigg]\,.
\label{t-xk}
\ea
The wave function \re{SoV-gen} has to diagonalize the transfer matrix,
Eq.~\re{ei-tu}. Taking into account \re{t-xk}, we find that the wave function in
the separated coordinates satisfies a multi-dimensional Baxter equation
\be
t_N(x_k)\,\Phi_{\mybf{q}}(\Mybf{x})=(x_k+is)^N\,\Phi_{\mybf{q}}
(\Mybf{x}+i\Mybf{e}_k)+ (x_k-is)^N\,\Phi_{\mybf{q}}(\Mybf{x}-i\Mybf{e}_k)\,,
\label{Bax-multi}
\ee
with $k=1,\ldots,N-1$. Here we assumed that the integration contour in
\re{SoV-gen} can be shifted into the complex $\Mybf{x}-$plane. In addition, we
took into account that the measure $\mu(\Mybf{x})$ satisfies the following
finite-difference equation~\cite{Sklyanin}
\be
\frac{\mu(\Mybf{x}+i\Mybf{e_k})}{\mu(\Mybf{x})}=\frac{\Delta_+(x_k)}{\Delta_-(x_k+i)}
\prod_{j\neq k}\frac{x_k-x_j+i}{x_k-x_j}\,,
\label{measure-rec}
\ee
with $\Delta_\pm(x_k)$ defined in \re{coef-ad}. This relation follows from the
condition for $\widehat t_N(u)$ to be a self-adjoint operator in the SoV
representation, $\vev{p',\Mybf{x}'|\widehat
t_N(u)|p,\Mybf{x}}^*=\vev{p,\Mybf{x}|\widehat t_N(u^*)|p',\Mybf{x}'}$.

The same Baxter equation \re{Bax-multi} holds for a complex conjugated function
\be
\Phi^*_{\mybf{q}}(\Mybf{x})\,\delta(p-p') =
\vev{\Psi_{\mybf{q},p}|p',\Mybf{x}}\,.
\label{conj-fun}
\ee
To see this one uses \re{ei-tu} together with hermiticity of the transfer matrix
\be t_N(x_k)\vev{\Psi_{\mybf{q},p}|p',\Mybf{x}}=
\vev{\widehat t_N(x_k)\Psi_{\mybf{q},p}|p',\Mybf{x}}=
\vev{\Psi_{\mybf{q},p}|(A_N(x_k)+D_N(x_k)|p',\Mybf{x}}\,.
\ee
Here in the last relation $A_N(x_k)$ and $D_N(x_k)$ act on $\ket{p',\Mybf{x}}$ as
shift operators, Eq.~\re{shiftD}. Using \re{conj-fun}, one finds that
$\Phi^*_{\mybf{q}}(\Mybf{x})$ satisfies \re{Bax-multi} and, therefore, one should
expect that $\Phi^*_{\mybf{q}}(\Mybf{x})\sim\Phi_{\mybf{q}}(\Mybf{x})$ for real
$\Mybf{x}$. We shall verify this relation below (see Eqs.~\re{facQ} and
\re{Q-conj}).

As was mentioned in the Introduction, Eqs.~\re{measure-rec} and \re{Bax-multi} do
not allow us to uniquely determine the integration measure and the eigenfunctions
in the SoV representation since their solutions are defined up to a
multiplication by an arbitrary periodic function $f(x)=f(x\pm i)$. To fix the
ambiguity one has to impose the additional conditions on the solutions to
\re{measure-rec} and \re{Bax-multi}. As we will show in the next Section, these
conditions follow from the explicit expression for the kernel of the unitary
transformation to the SoV representation, $U_{p,\mybf{x}}(z)$.

\subsection{Construction of the SoV representation}
\label{SOV}

\begin{figure}[t]
\psfrag{w2}[cc][cc]{$\bar w_2$}
\psfrag{w3}[cc][cc]{$\bar w_3$}
\psfrag{w5}[cc][cc]{$\bar w_N$}
\psfrag{w4}[cc][cc]{$\bar w_{N-1}$}
\psfrag{z1}[cc][cc]{$z_1$}
\psfrag{z2}[cc][cc]{$z_2$}
\psfrag{z5}[cc][cc]{$z_3$}
\psfrag{z4}[cc][cc]{$z_N$}
\psfrag{z3}[cc][cc]{$z_{N-1}$}
\psfrag{s1}[cr][cc]{$s-iu$}
\psfrag{s2}[cr][cc]{$s+iu$}

\centerline{\epsfxsize14.0cm\epsfbox{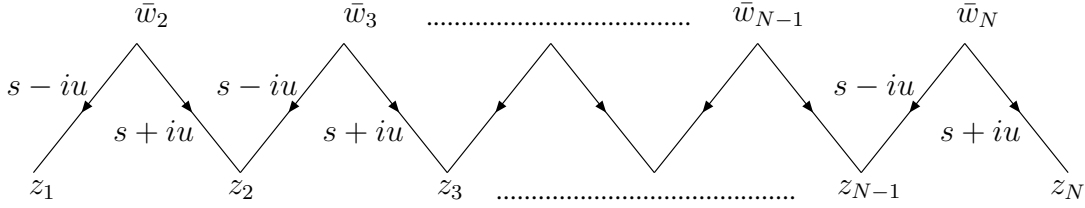}}
\vspace*{0.5cm}
\caption[]{Diagrammatical representation of the function
$\Lambda_u(z_1,\ldots,z_N|w_2,\ldots,w_N)$. The arrow with the index $\alpha$
that connects the points $\bar w$ and $z$ stands for $(z-\bar w)^{-\alpha}$.}
\label{fig1}
\end{figure}

To determine the kernel $U_{p,\mybf{x}}(z)$ one has to solve the system \re{BxN}
under the additional condition \re{Weyl}. Constructing solution for
$U_{p,\mybf{x}}(z)$, we shall follow the approach developed in Ref.~\cite{DKM} in
application to the quantum $SL(2,\mathbb{C})$ spin magnet.

To begin with, one considers the first equation in the system \re{BxN}
\be\label{step-1}
B_N(x_1)U_{p,\mybf{x}}(z_1,\ldots,z_N)=0\,.
\ee
Its solution can be found by making use of the invariance of the transfer matrix
\re{tu} under local gauge transformations of the Lax operators \re{Lax} and the
monodromy matrix \re{Tu}~\cite{PG,SD}
\begin{equation}
L_k(u)\to \widetilde L_{k}(u) = M_k^{-1}\,L_k(u)\,M_{k+1}\,,\qquad
{\mathbb T}_N(u) \to
\widetilde {\mathbb T}_N(u) = M_1^{-1}\, {\mathbb T}_N(u)\, M_1\,.
\label{rot}
\end{equation}
Here $M_k$ are arbitrary $2\times2$ matrices, such that $M_{N+1}=M_{1}$ and
$\det M_k\neq 0$. Let us choose the matrices $M_k$ as
\begin{equation}
\label{S-trick}
M_k=\left(\begin{array}{cc}1& \bar w_k^{\,-1}\\0&1\end{array}\right)\,,\qquad
M_k^{-1}=\matrix{1& -\bar w_k^{\,-1}\\0 &1}
\end{equation}
with $\bar w_1,\ldots,\,\bar w_N$ being arbitrary gauge parameters. It is
straightforward to verify that being applied to the function
\begin{equation}
\phi_u(z_k;\bar w_k,\bar w_{k+1})=(z_k-\bar w_k)^{-s-iu}\,(z_k-\bar
w_{k+1})^{-s+iu}\,,
\label{bxk}
\end{equation}
the gauge rotated Lax operator $\widetilde L_{k}(u)$ takes the form of a lower
triangle matrix
\be
\widetilde L_{k}(u) \cdot \phi_u(z_k;\bar w_k,\bar w_{k+1})=
\lr{\begin{array}{cc}
  (u+is)\phi_{u+i} & 0 \\
  \ast & (u-is)\phi_{u-i}
\end{array}}\,.
\label{L-aux}
\ee
This suggests to define the following function
\begin{equation}\label{Bxal}
Y_{u}(z,\bar w)=\prod_{k=1}^N\phi_u(z_k;\bar w_k,\bar w_{k+1})=
\prod_{k=1}^N(z_k-\bar w_k)^{-s-iu}\,(z_k-\bar w_{k+1})^{-s+iu}\,.
\end{equation}
Aside from the coordinates $z=(z_1,\ldots,z_N)$, it depends on auxiliary
variables $\bar w=(\bar w_1,\ldots,\bar w_N)$ and the spectral parameter $u$.
Denoting the matrix elements of $\widetilde {\mathbb T}_N(u)$ as $\widetilde
A_N(u),\ldots,\widetilde D_N(u)$ in the same manner as \re{Tu}, one finds from
\re{L-aux} that $\widetilde B_N(u) Y_{u}(z,\bar w)=0$ and
\be
\widetilde A_N(u) Y_{u}(z,\bar w)= (u+is)^N Y_{u+i}(z,\bar w)\,,\qquad
\widetilde D_N(u) Y_{u}(z,\bar w)= (u-is)^N Y_{u-i}(z,\bar w)\,.
\label{semi-bax}
\ee
As follows from \re{rot}, the operators $\widetilde A_N(u),\ldots,\widetilde
D_N(u)$ are given by linear combinations of the operators $A_N(u),\ldots,D_N(u)$
with the coefficients depending on $1/\bar w_1$. For $\bar w_1\to\infty$ the two
sets of the operators coincide since $M_1=\II$ and $\widetilde {\mathbb T}_N(u)=
{\mathbb T}_N(u)$. In this limit, the r.h.s.\ of \re{Bxal} scales as
$Y_{u}(z,\bar w)\sim \bar w_1^{2s}\, \Lambda_u(z,\bar w)$ with
\ba
&&\Lambda_u(z_1,\ldots,z_N|\bar w_2,\ldots,\bar w_N)=
\nonumber
\\
&&\qqqquad (z_1-\bar w_2)^{-s+iu}\left(
\prod_{k=2}^{N-1}(z_k-\bar w_k)^{-s-iu}(z_k-\bar w_{k+1})^{-s+iu}\right)
(z_N-\bar w_N)^{-s-iu}\,.
\label{La}
\ea
One deduces from \re{semi-bax} that for arbitrary $\bar w_2,\ldots,\bar w_N$ the
function $\Lambda_{u}(z,\bar w)$ satisfies the relations
\ba
&&
B_N(u) \Lambda_{u}(z,\bar w)=0\,,
\nonumber
\\
&&
A_N(u) \Lambda_{u}(z,\bar w)= (u+is)^N \Lambda_{u+i}(z,\bar w)\,,
\label{Lam-rel}
\\
&&
D_N(u)
\Lambda_{u}(z,\bar w)= (u-is)^N \Lambda_{u-i}(z,\bar w)\,.
\nonumber
\ea%
\begin{figure}[t]
\psfrag{z1}[cc][cc]{$z_1$}
\psfrag{z2}[cc][cc]{$z_2$}
\psfrag{z3}[cc][cc]{$z_{N-1}$}
\psfrag{z4}[cl][cc]{$z_N$}
\psfrag{v1}[cc][cc]{$\bar v_3$}
\psfrag{v3}[cc][cc]{$\bar v_N$}
\psfrag{x1}[cr][cc]{$s-ix_1$}
\psfrag{x2}[cr][cc]{$s+ix_1$}
\psfrag{y1}[cr][cc]{$s-ix_2$}
\psfrag{y2}[cr][cc]{$s+ix_2$}
\psfrag{f1}[rt][cc][1][90] {$\alpha$}
\psfrag{f2}[cb][cb][1][90]{$-\alpha$}
\psfrag{f1}[cc][cc]{$\alpha$}
\psfrag{f2}[rr][cc]{$-\alpha$}
\centerline{\epsfxsize15.0cm\epsfbox{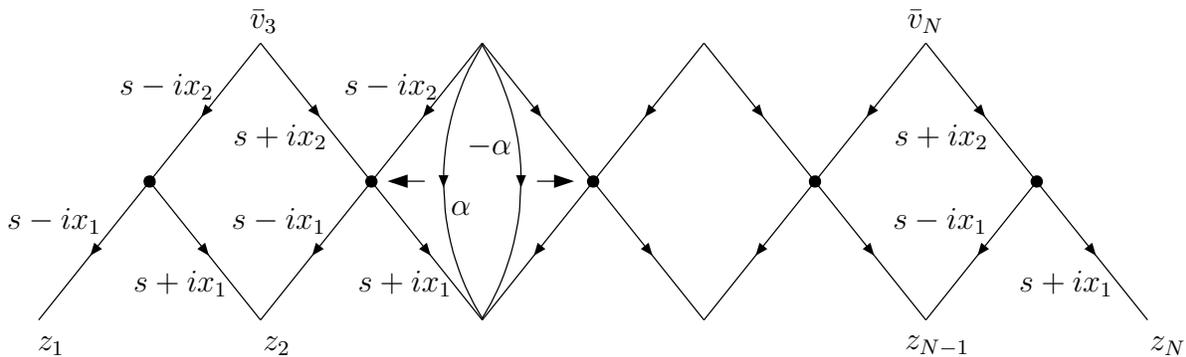}}
\vspace*{0.5cm}
\caption[]{Diagrammatical proof of the identity \re{L-perm}. One inserts
two vertical lines with the indices $\pm\alpha=\pm i(x_2-x_1)$ into one of the
rhombuses and displaces them in the directions indicated by arrows with a help of
identities shown in Figures~\ref{comm-f} and \ref{amp1}.}
\label{LL}
\end{figure}%
%
It proves extremely useful to translate the obtained expression for the
$\Lambda-$function, Eq.~\re{La}, into a language of Feynman diagrams~\cite{DKM}.
Namely, one represents each factor $(z-\bar w)^{-\alpha}$ in the r.h.s.\ of
\re{La} by an arrow with the index $\alpha$ that starts in the point $\bar w$ and
ends in $z$. In this way, the $\Lambda-$function can be represented as a Feynman
diagram shown in Figure~\ref{fig1}.

Making use of the first relation in \re{Lam-rel}, one finds that the general
solution to Eq.~\re{step-1} can be written as a convolution of the
$\Lambda-$function with an arbitrary weight function $Z_{N-1}(w_2,\ldots,w_N)$,
which is holomorphic in the upper half-plane and depends, in general, on the
separated variables $x_2,\ldots,x_N$
\ba
U_{p,\mybf{x}}(z_1,\ldots,z_N)&=& \int \mathcal{D}^{N-1}w \,
\Lambda_{x_1}(z_1,\ldots,z_N|\bar w_2,\ldots,\bar w_N)\,Z_{N-1}(w_2,\ldots,w_N)
\nonumber
\\
&\equiv& [\Lambda_N(x_1)\otimes Z_{N-1}](z_1,\ldots,z_N)\,,
\label{conv}
\ea
where $\mathcal{D}^{N-1}w=\mathcal{D}w_2\ldots \mathcal{D}w_N$ with $\bar
w_n=w_n^*$. Here 
the notation was introduced for the integral operator $\Lambda_N(x_1)$ with the
kernel defined by the function $\Lambda_{x_1}(z_1,\ldots,z_N|\bar w_2,\ldots,\bar
w_N)$.

Let us now require that the function $U_{p,\mybf{x}}(z)$ defined in \re{conv} has
to satisfy the second equation from the system \re{BxN},
$B_N(x_2)\,U_{p,\mybf{x}}(z)=0$. This leads to a rather complicated equation for
the weight function $Z_{N-1}(w)$. To solve this equation, we shall propose a
particular ansatz for $Z_{N-1}(w)$ and verify that it leads to the expression for
$U_{p,\mybf{x}}(z)$ which obeys \re{BxN}. Suppose that there exists such
$Z_{N-1}(w)$ that the resulting expression for \re{conv} is a symmetric function
of $x_1$ and $x_2$. Then, the above requirement will be automatically fulfilled
in virtue of \re{step-1}. The definition of $Z_{N-1}(w)$ is based on the
following remarkable identity
\be
[\,\Lambda_N(x_1)\otimes\Lambda_{N-1}(x_2)\,](z_1,\ldots,z_N|v_3,\ldots,v_N)=
[\,\Lambda_N(x_2)\otimes\Lambda_{N-1}(x_1)\,](z_1,\ldots,z_N|v_3,\ldots,v_N)\,,
\label{L-perm}
\ee
where the operator $\Lambda_{N}(x_2)$ was defined for arbitrary $N$ in \re{conv}
and \re{La}. The both sides of this relation can be rewritten as a convolution of
two $\Lambda-$functions, Eq.~\re{La}. In the Feynman diagram representation, the
l.h.s.\ of \re{L-perm} is described by the diagram shown in Figure~\ref{LL}.
Having the identity \re{L-perm} in mind, we choose the function $Z_{N-1}(w)$ in
the following form
\be
Z_{N-1}(w_2,\ldots,w_N)=\left[\Lambda_{N-1}(x_2)\otimes
Z_{N-2}\right](w_2,\ldots,w_N)\,,
\label{ZL}
\ee
where the function $Z_{N-2}=Z_{N-2}(v_3,\ldots,v_N)$ is holomorphic in the upper
half-plane, $\Im v_k>0$, and depends, in general, on $x_3,\ldots,x_N$.
Substituting \re{ZL} into \re{conv} and making use of \re{L-perm} we find that
$U_{p,\mybf{x}}(z)$ is invariant under permutations of $x_1$ and $x_2$

The proof of \re{L-perm} can be performed diagrammatically without doing any
calculation. It is based on elementary permutation identities shown
diagrammatically in Figures~\ref{comm-f} and \ref{amp1}. Their derivation can be
found in Appendix~\ref{A}. Similar identities have been also found for the Toda
model~\cite{PG} and the $SL(2,\mathbb{C})$ spin chain~\cite{DKM}. To prove
\re{L-perm} one inserts two auxiliary lines with the indices $\pm i(x_1-x_2)$
into one of the rhombuses in Figure~\ref{LL}. Since the lines are attached to the
same points and the sum of their indices equals zero, this transformation does
not affect the l.h.s.\ of \re{L-perm}. Then, one applies the permutation identity
shown in Figure~\ref{comm-f} and moves the line with the index $i(x_2-x_1)$ to
the left part of the diagram until it reaches the leftmost rhombus in which case
one applies the identity shown in Figure~\ref{amp1}. Performing similar
transformations on the second line with the indices $-i(x_2-x_1)$, one moves it
to the right part of the diagram. In this way, one obtains the initial Feynman
diagram but with the variables $x_1$ and $x_2$ interchanged, thus proving
\re{L-perm}.

\begin{figure}[t]
\psfrag{z1}[cc][cc]{$z_1$}
\psfrag{z2}[cc][cc]{$z_2$}
\psfrag{z3}[cc][cc]{$z_{N-1}$}
\psfrag{z4}[cl][cc]{$z_N$}
\psfrag{w1}[cc][cc]{$\bar w_N$}
\psfrag{x1}[cr][cc]{$\alpha_1$}
\psfrag{x2}[cr][cc]{$\alpha_2$}
\psfrag{y1}[cr][cc]{$\beta_1$}
\psfrag{y2}[cc][cc]{$\beta_2$}

\psfrag{x3}[cr][cc]{$\alpha_{N-1}$}
\psfrag{y3}[cl][cc]{$\beta_{N-1}$}

\centerline{\epsfxsize10.0cm\epsfbox{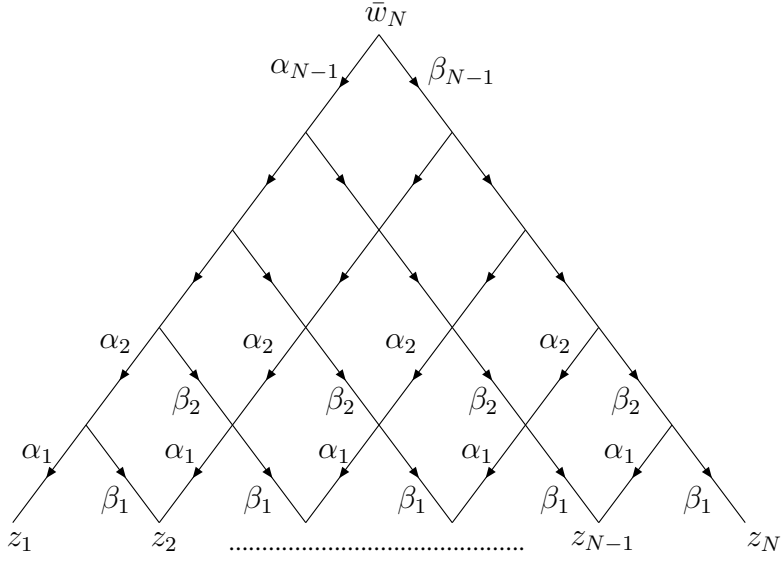}}
\vspace*{0.5cm}
\caption[]{Diagrammatical representation of the function $U_{\mybf{x}}(z;\bar w_N)$.
The indices $\alpha_k=s-ix_k$ and $\beta_k=s+ix_k$ parameterize the corresponding
factors entering \re{La}. The $SL(2,\mathbb{R})$ integration \re{measure} over
the position of internal vertices is implied.}
\label{Pyramid}
\end{figure}

It is now 
straightforward to write a general expression for the function
$U_{p,\mybf{x}}(z)$ satisfying the relations \re{BxN} and \re{Weyl}
\be
U_{p,\mybf{x}}(z_1,\ldots,z_N)= p^{Ns-1/2}
\int_{\Im w_N >0} \mathcal{D}w_N\, \e^{ip\,w_N}\, U_{\mybf{x}}(\vec z\,;\bar w_N)\,,
\label{B-ei}
\ee
where $\vec z=(z_1,\ldots,z_N)$ and $U_{\mybf{x}}(\vec z\,;\bar w_N)$ is
factorized into the product of $N-1$ operators
\be
U_{\mybf{x}}(\vec z\,;\bar w_N)=
\left[\Lambda_N(x_1)\otimes\Lambda_{N-1}(x_2)\otimes\ldots\otimes
\Lambda_2(x_{N-1})
\right](z_1,\ldots,z_N|\bar w_N)\,.
\label{U}
\ee
Here the operator $\Lambda_n(x_{N+1-n})$ has an integral kernel
$\Lambda_{x_{N+1-n}}(z_1,...,z_n|\bar w_2,...,\bar w_n)$
 defined in \re{La}. It depends on a single separated
coordinate $x_{N+1-n}$ (with $n=2,\ldots,N$) and being applied to an arbitrary
function of $w_2,\ldots,w_n$ it increases the number of its argument by 1 (see
Eq.~\re{conv}). The expression for the function $U_{\mybf{x}}(\vec z\,;\bar
w_N)$, Eq.~\re{U}, admits a simple diagrammatic representation. Replacing each
$\Lambda-$operator in \re{U} by the corresponding Feynman diagram (see
Figure~\ref{fig1}) one obtains that $U_{\mybf{x}}(\vec z\,;\bar w_N)$ can be
represented as the ``Pyramide du Louvre'' diagram shown in Figure~\ref{Pyramid}.
It consists of $(N-1)-$rows with each row denoting a single $\Lambda-$operator in
\re{U}. The $k-${th} row of the pyramid consists of $(N-k)-$lines carrying the
index $\alpha_k=s-ix_k$ and the same number of lines with the index
$\beta_k=s+ix_k$ depending on the separated variable $x_k$. Remarkably enough, a
pyramid diagram similar to the one shown in Figure~\ref{Pyramid} describes the
SoV transformation for the quantum $SL(2,\mathbb{C})$ spin chain~\cite{DKM,VL}
and the periodic Toda chain~\cite{KL}. We would like to stress that a
pyramid-like form of $U_{\mybf{x}}(\vec z\,;\bar w_N)$ is a consequence of the
factorized form of the kernel \re{U}, which is expected to be a general feature
of the $SL(2)$ magnets and related quantum integrable models.

Let us verify that the obtained expression for the function $U_{\mybf{x}}(\vec
z\,;\bar w_N)$, Eq.~\re{U}, verifies the defining relations \re{BxN} and
\re{Weyl}. Indeed, it follows from Eqs.~\re{L-perm} and \re{step-1} that
$U_{\mybf{x}}(\vec z\,;\bar w_N)$ is a completely symmetric function of
$x_1,\ldots,x_{N-1}$, which is nullified by the operators $B_N(x_k)$ for
$k=1,\ldots,N-1$. The integration over $w_N$ in \re{B-ei} ensures that
$U_{p,\mybf{x}}(z)$ is an eigenfunction of the operator of the total momentum
$iS_{-}$. To see this, one has to take into account that the kernel of the
$\Lambda-$operator, Eq.~\re{La}, depends on differences of the coordinates and,
therefore, the function $U_{\mybf{x}}(\vec z\,;\bar w_N)$ is translation
invariant, $U_{\mybf{x}}(z+\epsilon;\bar w_N+\epsilon)= U_{\mybf{x}}(z;\bar w_N)$
with $\epsilon$ real.%
\footnote{Notice that the $SL(2,\mathbb{R})$ integration in Eq.~(\ref{B-ei})
can be replaced by the integration over the real axis $\int_{-\infty}^\infty d
w_N$. According to \re{mom-rep}, the result will differ by the normalization
factor $p^{1-2s}\Gamma(2s)$.} The normalization factor $p^{Ns-1/2}$ was inserted
in the r.h.s.\ of \re{B-ei} for the later convenience.

By the construction, $U_{\mybf{x}}(\vec z;\bar w_N)$ is a holomorphic function of
$\vec z$ and $w_N=\bar w_N^*$ in the upper-half plane. One finds from \re{B-ei}
that $U_{p,\mybf{x}}(z)$ vanishes for $p<0$ since the integration contour over
$\Re w_N$ can be closed into the lower half-plane. Then, it follows from
\re{SoV-gen} that the eigenfunctions $\Psi_{\mybf{q},p}\,(z)$ are different from
zero only for $p>0$. This is a common feature of states belonging to the Hilbert
space \re{scalar} and \re{norm}. The eigenstates with negative momenta, $p<0$,
can be constructed on the space of functions holomorphic in the lower-half plane.

Since the kernel of the $\Lambda-$operator satisfies \re{Lam-rel}, one verifies
that, in agreement with \re{shiftD}, the same relations hold for the function
$U_{p,\mybf{x}}(z)$. Using \re{tu} one finds
\be
\widehat t_N(x_k)\,U_{p,\mybf{x}}(z)=(x_k+is)^N\,
U_{p,\mybf{x}+i\mybf{e}_k}(z)+(x_k-is)^N\, U_{p,\mybf{x}-i\mybf{e}_k}(z)\,.
\ee
As was shown in Section~3.1, this leads to the multi-dimensional Baxter equation
for the wave function in the SoV representation, Eq.~\re{Bax-multi}. It follows
from \re{B-ei} and \re{U} that $U_{p,\mybf{x}}(z)$ is an entire function of
$x_1,\ldots,x_{N-1}$ and, as a consequence, the solutions to \re{Bax-multi} have
to possess the same property. Indeed, the analytical properties of
$U_{p,\mybf{x}}(z)$ are in one-to-one correspondence with possible divergences of
the Feynman integral in the r.h.s.\ of \re{U}. Using \re{La} one verifies that
the integral is convergent for arbitrary complex $\Mybf{x}$.

\subsection{Recurrence relations/Contour integral representation}

Expressions \re{B-ei} and \re{U} have a simple recursive form as a function of
the number of particles $N$. Increasing this number, $N\to N+1$, one has just to
add an additional row to the pyramid diagram in Figure~\ref{Pyramid}. Namely
\ba
U_{x_1,\ldots,x_{N-1}}(z_1,\ldots,z_N;\bar w_N)&=&
\nonumber\\
&&\hspace*{-40mm}\int \mathcal{D}^{N-1}v \,
\Lambda_{x_{N-1}}(z_1,\ldots,z_N|\bar v_1,\ldots,\bar v_{N-1})\,
 U_{x_1,\ldots,x_{N-2}}(v_1,\ldots,v_{N-1};\bar w_N)\,,
\ea
with the $\Lambda-$function defined in \re{La}.
 The $SL(2,\mathbb{R})$ integrals
over the $v-$coordinates can be simplified by making use of the identity
\re{drift}. In this way, one gets
\ba
&& U_{x_1,\ldots,x_{N-1}}(z_1,\ldots,z_N;\bar
w_N)=\left[i^{2s}B(s+ix_{N-1},s-ix_{N-1})\right]^{-N+1}
\nonumber
\\
&&\quad
\times\prod_{k=1}^{N-1} \int_0^1 d\tau_k\,
\tau_k^{s-ix_{N-1}-1}(1-\tau_k)^{s+ix_{N-1}-1}
U_{x_1,\ldots,x_{N-2}}(z_1(\tau_1),\ldots,z_{N-1}(\tau_{N-1});\bar w_N)\,,
\label{rec-rel}
\ea
where $B(x,y)$ is the Euler beta-function and $z_k(\tau_k)=(1-\tau_k)z_k+\tau_k
z_{k+1}$, so that $z_k(0)=z_{k}$ and $z_k(1)=z_{k+1}$.
The $U-$function in the r.h.s.\ is described by the
same pyramid diagram but with one row less. Its end-points are located in between
the end-points of the pyramid in the l.h.s.\ of \re{rec-rel}. At $N=2$ one gets
from \re{U}
\be
U_{x_1}(z_1,z_2;\bar w_2)=(z_1-\bar w_2)^{-s+ix_1} (z_2-\bar w_2)^{-s-ix_1}\,.
\label{N=2}
\ee
Repeatedly applying \re{rec-rel} and using \re{N=2} as a boundary condition, one
can express $U_{\mybf{x}}(\vec z\,;\bar w_N)$ as a product of one-dimensional
nested contour integrals. To save space, we do not present here the explicit
expression.

\section{Integration measure}
\label{meas}

To calculate the integration measure in the SoV representation, $\mu(\Mybf{x})$,
one has to substitute the obtained expressions for the functions
$U_{p,\mybf{x}}(z_1,\ldots,z_N)$, Eq.~\re{B-ei}, into the orthogonality condition
\re{U-ort} and perform integration. In spite of the fact that a multi-dimensional
integral in \re{U-ort} seems to be rather complicated, it can be easily evaluated
by making use of the diagrammatical representation for $U_{p,\mybf{x}}(z)$ (see
Figure~\ref{Pyramid}). The analysis goes along the same lines as calculation of
the integration measure for the $SL(2,\mathbb{C})$ magnet~\cite{DKM}.

Let us substitute \re{B-ei} into \re{U-ort} and consider the following scalar
product
\be
\vev{w_N',\Mybf{x}'|w_N,\Mybf{x}}=\int \mathcal{D}^N z \,
U_{\mybf{x}}(z_1,...,z_N;\bar w_N) \lr{U_{\mybf{x'}}(z_1,...,z_N;\bar w_N')}^*\,.
\label{vev-aux}
\ee
To evaluate \re{U-ort} one has to perform a Fourier transformation of
$\vev{w_N',\Mybf{x}'|w_N,\Mybf{x}}$ with respect to $w_N$ and $w_N'$ and multiply
it by the additional factor $p^{2Ns-1}$. The function $U_{\mybf{x}}(\vec z\,;\bar
w_N)$ entering \re{vev-aux} is represented by the pyramid diagram shown in
Figure~\ref{Pyramid}. To obtain $\lr{U_{\mybf{x}}(\vec z\,;\bar w_N')}^*$ one has
to replace in Eqs.~\re{B-ei} and \re{La} the holomorphic ``propagators'' $(z-\bar
w)^{-\alpha}$ by complex conjugated expressions ${(\bar
z-w)^{-\alpha^*}}=\e^{i\pi\alpha^*}{(w-\bar z)^{-\alpha^*}}$. The function
$\lr{U_{\mybf{x}}(z;\bar w_N')}^*$ can be represented by the same pyramid diagram
if one replaces in Figure~\ref{Pyramid} the indices $\alpha_k$ and $\beta_k$ by
their conjugated expressions, $\alpha_k^*=\beta_k$ and $\beta_k^*=\alpha_k$,
respectively, and flips the direction of all arrows. Each arrow is accompanied by
the additional factor $\e^{i\pi\alpha_k^*}$ or $\e^{i\pi\beta_k^*}$. Combining
these factors together along the rows of the pyramid diagram and using the
identity $\alpha_k+\beta_k=2s$, one finds that their total product equals
$\e^{i\pi N(N-1)s}$. As we will show below, in the final expression for the
measure this factor cancels against a similar factor coming from the
$z-$integration in \re{vev-aux}.

\begin{figure}[th]
\psfrag{p1}[cr][cc]{$s-ix_1$}
\psfrag{p2}[cl][cc]{$s+ix_1$}
\psfrag{p3}[cr][cc]{$s+ix_1'$}
\psfrag{p4}[cl][cc]{$s-ix_1'$}
\psfrag{w2}[cc][cc]{$\bar w_2$}
\psfrag{w1}[cc][cc]{$w_2'$}
\psfrag{z1}[cr][cc]{$z_1$}
\psfrag{z2}[cl][cc]{$z_2$}
\centerline{\epsfxsize3.5cm\epsfbox{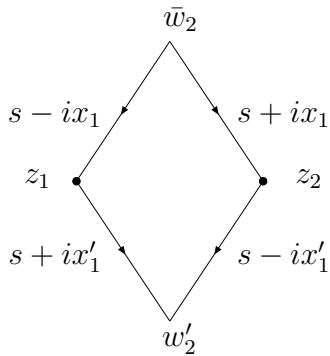}}
\vspace*{0.5cm}
\caption[]{The scalar product of two pyramid diagrams at $N=2$.}
\label{proof2}
\end{figure}

It is convenient to flip horizontally the conjugated pyramid diagram, so that the
point $w_N'$ will be located at the bottom of the diagram and the points
$z_1,\ldots,z_N$ at the top. The scalar product \re{vev-aux} is obtained by
sewing together the pyramid and its conjugated counterpart at the points
$z_1,\ldots,z_N$. The resulting Feynman diagram takes the form of a big rhombus
built out of $(N-1)^2-$elementary rhombuses (see the leftmost diagram in
Figure~\ref{proof} below). Its tips have the coordinates $\bar w_N$ and $w_N'$.
Notice that the indices $\alpha_k$ and $\beta_k$ in the upper and lower part of
this rhombus depend on two different sets of the separated coordinates $\Mybf{x}$
and $\Mybf{x}'$, respectively.

\begin{figure}[th]
\psfrag{a}[cc][cc]{$(a)$}
\psfrag{b}[cc][cc]{$(b)$}
\psfrag{c}[cc][cc]{$(c)$}
\psfrag{d}[cc][cc]{$(d)$}
\psfrag{e}[cc][cc]{$(e)$}
\centerline{\epsfxsize14.0cm\epsfbox{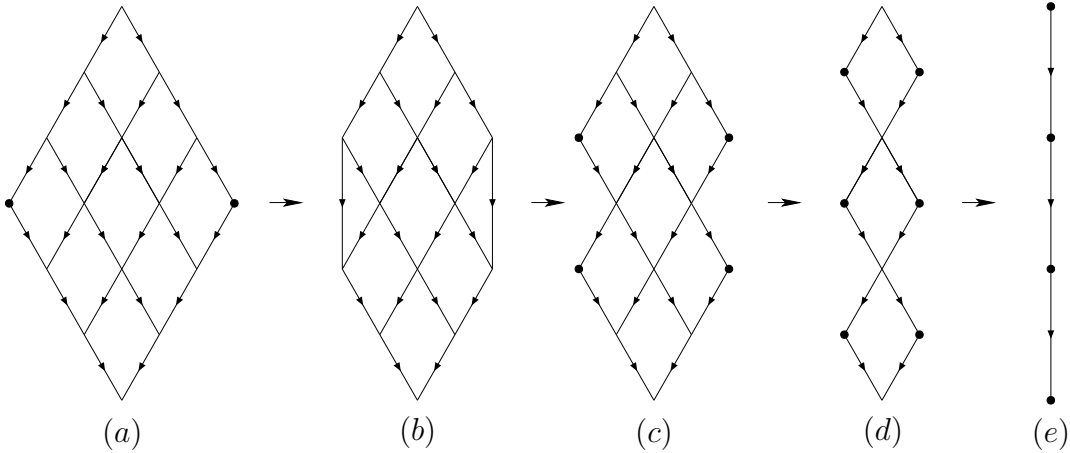}}
\vspace*{0.5cm}
\caption[]{ Diagrammatical  calculation of the scalar product of two
pyramids. The fat points indicate the vertices which can be integrated out using
the chain relation shown in Figure~\ref{Chain}.}
\label{proof}
\end{figure}

Let us first calculate \re{vev-aux} at $N=2$. The corresponding rhombus diagram
is shown in Figure~\ref{proof2}. Integration over $z_1$ and $z_2$ can be easily
performed using the chain relation shown in Figure~\ref{Chain} (see Appendix A
for details) leading to $\vev{w_2',x_1'|w_2,x_1}\sim (w_2'-\bar w_2)^0$. Its
Fourier transformation with respect to $w_2$ and $w_2'$ leads to the expression
for $\vev{p',x_1'|p,x_1}$, which is divergent at $p=p'$. This means that, in
agreement with \re{U-ort}, $\vev{p',x_1'|p,x_1}$ should be understood as a
distribution. To identify its form one has to regularize the corresponding
Feynman integrals. As was shown in \cite{DKM}, this can be achieved by shifting
the indices as
\be
\alpha_k\to \alpha_k+\epsilon\,,\qquad \beta_k\to\beta_k+\epsilon
\label{shift}
\ee
and carefully examining the limit $\epsilon\to 0$ 
\be
\vev{p',\Mybf{x'}|p,\Mybf{x}}=(pp')^{Ns-1/2}
\lim_{\epsilon\to 0}\int \mathcal{D}^N w_N \e^{ip\,w_N}\int
\mathcal{D}^N w_N'\e^{-ip'\,\bar w_N}
\vev{w_N',\Mybf{x}'|w_N,\Mybf{x}}_\epsilon\,.
\label{limit}
\ee
Repeating the calculation of the rhombus diagram in Figure~\ref{proof2}, one gets
\be
\vev{w_2',x_1'|w_2,x_1}_\epsilon=\e^{2i\pi s}\gamma_\epsilon(x_1,x_1')\cdot
(w_2'-\bar w_2)^{-4\epsilon}\,, \label{vev-2}
\ee
where the notation was introduced for (see Eq.~\re{As})
\ba
\gamma_\epsilon(x_1,x_1')&=&a\lr{s-ix_1+\epsilon,s+ix_1'+\epsilon}\,
a\lr{s+ix_1+\epsilon,s-ix_1'+\epsilon}
\nonumber\\
&=&\frac{\e^{-2i\pi
s}\Gamma(2\epsilon+i(x_1-x_1'))\Gamma(2\epsilon+i(x_1'-x_1))\Gamma^2(2s)}{
\Gamma(\epsilon+s+ix_1)\Gamma(\epsilon+s-ix_1)\Gamma(\epsilon+s+ix_1')\Gamma(\epsilon+s-ix_1')}\,.
\label{factor1}
\ea
Then, one substitutes \re{vev-2} into \re{limit}, performs its Fourier
transformation with a help of \re{p-rep} and applies the identity
\be
\lim_{\epsilon\to
0}\frac{\Gamma(2\epsilon-ix)\Gamma(2\epsilon+ix)}{\Gamma(4\epsilon)}=
2\pi\,\delta(x)\,.
\label{reg-delta}
\ee
to get the expression for $\vev{p',x_1'|p,x_1}\sim\delta(p-p')\delta(x_1-x_1')$,
which matches the r.h.s.\ of Eq.~\re{U-ort} at $N=2$ for
\be
\mu(x_1)=\frac{1}{2\pi}
\left[\frac{\Gamma(s+ix_1)\Gamma(s-ix_1)}{\Gamma^2(2s)}\right]^2\,.
\label{measure_2}
\ee
This expression defines the integration measure in the SoV representation at
$N=2$. It is interesting to note that $\mu(x_1)$ coincides with the weight
function for the continuous Hanh polynomials~\cite{Koekoek}.

The calculation of the scalar product \re{vev-aux} for $N\ge 3$ is shown
schematically in Figure~\ref{proof}. At the first step, one uses the chain
relation, Figure~\ref{Chain}, to integrate out the left- and rightmost vertices
with the coordinates $z_1$ and $z_N$, respectively. This transformation replaces
two pairs of lines by two vertical lines with the indices $\pm i(x_1-x_1')$ (see
Figure~\ref{proof}b) and brings the factor $\gamma_{\epsilon=0}(x_1,x_1')$
defined in \re{factor1}. Repeatedly applying the permutation identity,
Figure~\ref{comm-f}, we move one of the vertical lines horizontally through the
diagram in the direction of another one until they meet and annihilate each
other. In this way, we arrive at the diagram shown in Figure~\ref{proof}c. In
comparison with Figure~\ref{proof}a, it has two vertices less and the parameters
$x_1$ and $x_1'$ are interchanged. In this diagram there are already four
vertices which can be integrated out using the chain relation,
Figure~\ref{Chain}. Further simplification amounts to repetition of the steps
just described. At each next step the number of vertices in the diagram is
reduced and the $\Mybf{x}-$parameters got interchanged, $x_k\leftrightarrow
x_{\!j}^\prime$. Continuing this procedure, one obtains the diagram shown in
Figure~\ref{proof}d, in which $(N-1)-$elementary rhombuses are aligned along the
vertical axis. It is accompanied by the factor $\prod_{j+k\le
N-1}\gamma_{\epsilon=0}(x_j,x_k')$ defined in \re{factor1}. Enumerating the
rhombuses in Figure~\ref{proof}d starting from the top, one finds that the $k$th
rhombus coincides with the $N=2$ diagram shown in Figure~\ref{proof2} upon
replacement $x_1$ and $x_1'$ by $x_{N-k}$ and $x_k'$, respectively. Based on the
$N=2$ calculations, one should expect that the chain of rhombuses produces the
contribution $\sim\prod_{k=1}^N\delta(x_{N-k}-x_k')$. This turns out to be
correct, but in order to see it one has to regularize the Feynman integrals
according to
\re{shift} and carefully examine their limit 
as $\epsilon\to 0$. For $\epsilon\neq 0$ each elementary rhombus in
Figure~\ref{proof}d can be replaced by a single line with the index
$4\varepsilon$ (see Figure~\ref{proof}e). This brings the additional factor
$\prod_{k=1}^{N-1}\gamma_\epsilon(x_k,x_{N-k}')$. Finally, one integrates out the
remaining $(N-2)-$vertices using the chain relation, Figure~\ref{Chain}, and
combines together all factors to find the following expression for the
regularized scalar product \re{vev-aux}
\ba
\vev{w_N',\Mybf{x}'|w_N,\Mybf{x}}_\epsilon&=&
\e^{i\pi N(N-1)s}\prod_{j+k\le N-1}\gamma_{\epsilon=0}(x_j,x_k')\prod_{k=1}^{N-1}
\gamma_\epsilon(x_k,x_{N-k}')
\nonumber\\
&\times&
 \e^{-i\pi(N-2)s}
\frac{\Gamma^{N-2}(2s)\Gamma(\Delta)}{\Gamma^{N-1}(4\epsilon)}
(w_N'-\bar w_N)^{-\Delta}\,,
\label{factor2}
\ea
where $\Delta=4\epsilon(N-1)-2s(N-2)$.

One substitutes \re{factor2} into \re{limit}, performs its Fourier transformation
with a help of \re{p-rep} and applies \re{reg-delta} to get after some algebra
\ba
\vev{p',\Mybf{x'}|p,\Mybf{x}}&=&(2\pi)^{N-1}\,
\,\delta(p-p')\,
\prod_{k=1}^{N-1}\delta(x_k-x_{N-k}')
\nonumber
\\
&\times& \Gamma^{N^2}(2s)\frac{\prod_{j<k}
\Gamma(i(x_k-x_j))\Gamma(-i(x_k-x_j)) }{\prod_{k=1}^{N-1}\left[\Gamma(s+ix_k)\Gamma(s-ix_k)\right]^N}\,.
\label{non-sym}
\ea
This relation agrees with Eq.~\re{U-ort}, but in distinction with the latter its
r.h.s.\ is not symmetric in $\Mybf{x}$. Indeed, simplifying the Feynman diagrams
in Figures~\ref{proof}, we have tacitly assumed that the
$\gamma_{\epsilon=0}-$factors entering the r.h.s.\ of \re{factor2} are finite
functions of $\Mybf{x}$ and $\Mybf{x'}$. Taking into account \re{factor1}, one
finds that this is true provided that $x_j\neq x_k'$ with $j,\,k=1,\ldots,N-1$
and $j+k\le N-1$.%
\footnote{To restore the missing terms in the r.h.s.\ of \re{non-sym}, one has
to use the symmetry of the pyramid in order to rearrange its rows and repeat the
same calculation.} Under these conditions, only one term survives in the r.h.s.\
of \re{U-ort} and it has the same form as \re{non-sym}.

Matching \re{non-sym} into \re{U-ort} one obtains the integration measure for
$N\ge 3$
\be\label{mfi}
\mu({\Mybf{x}})=c_N
\prod_{j,k=1\atop j<k}^{N-1}{(x_k-x_j)}\,\sinh(\pi(x_k-x_j))  \,
\prod_{k=1}^{N-1}\left[\Gamma(s+ix_k)\Gamma(s-ix_k)\right]^N\,,
\ee
with the normalization factor
$c_N=[\Gamma^{N^2}(2s)(2\pi)^{N-1}(N-1)!\pi^{(N-1)(N-2)/2}]^{-1}$. The following
comments are in order.

In the SoV representation, the measure \re{mfi} is a semi-positive definite
function on the space of real $\Mybf{x}-$variables. It vanishes at the
hyperplanes $x_j=x_k$. After analytical continuation to complex $\Mybf{x}$,
$\mu({\Mybf{x}})$ becomes a meromorphic function of $x_k$ $(k=1,\ldots,N-1$) with
the $N$th order poles located along the imaginary axis at $x_k=\pm i(s+n)$ with
$n\in\mathbb{N}$. The measure decreases exponentially fast when $x_k$ goes to
infinity along the real axis and the remaining $\Mybf{x}-$variables take finite
values
\be\label{m-lim}
\mu(\Mybf{x}) {\sim} \,\e^{-2\pi |x_k|}x_k^{2(Ns-1)}\,,
\ee
as  $x_k\to\infty$ and $\Im x_k={\rm fixed}$. One verifies that the obtained
expression for the measure, Eq.~\re{mfi}, satisfies the functional relation
\re{measure-rec}.

The transition function $U_{p,\mybf{x}}(z)$, Eqs.~\re{B-ei} and \re{U},
satisfies the completeness condition
\be
\int_0^\infty {dp}\int_{\mathbb{R}^{N-1}} d^{N-1}\Mybf{x}\,
\mu(\Mybf{x})\,
\left(U_{p,\mybf{x}}(w_1,\ldots,w_N)\right)^*\,U_{p,\mybf{x}}(z_1,\ldots,z_N)=
\mathbb{K}(z;w)\,,
\label{completeness}
\ee
where $\int d^{N-1}\Mybf{x}\equiv\prod_{k=1}^{N-1}\int_{-\infty}^\infty dx_k$ and
$\mathbb{K}(z;w)$ is the reproducing kernel (kernel of the unity operator) on the
quantum space of the model $\mathcal{V}_N$ (see Eq.~\re{K-ker})
\be
\left[\mathbb{K}\cdot\Psi\right](z_1,\ldots,z_N)
=\int \mathcal{D}^N w\,\prod_{k=1}^N\frac{\e^{i\pi s}}{(z_k-\bar
w_k)^{2s}}\Psi(w_1,\ldots,w_N)=\Psi(z_1,\ldots,z_N)\,,
\label{K}
\ee
with $\Psi(z_1,\ldots,z_N)$ holomorphic in the upper half-plane. The relation
\re{completeness} is verified at $N=2$ by an explicit calculation in
Appendix~\ref{A}. For $N\ge 3$ the calculation is more involved and will be
presented elsewhere.

\section{Eigenfunctions in the SoV representation}
\label{Wave}

Let us consider the properties of the eigenfunction in the SoV representation
$\Phi_{\mybf{q}}(\Mybf{x})$ defined in Eq.~\re{SoV-inv}. Using \re{B-ei},
\re{vev-aux} and \re{mom-rep}, one can rewrite \re{SoV-inv} as
\be
\vev{w_N,\Mybf{x}|\Psi_{\mybf{q},p}}=
\int \mathcal{D}^N z\,\lr{U_{\mybf{x}}(\vec z\,;\bar w_N)}^*\Psi_{\mybf{q},p}(\vec z)
=\e^{ipw_N} \Phi_{\mybf{q}}(\Mybf{x})\cdot \frac{\theta(p)\,
p^{-s(N-2)-1/2}}{\Gamma(2s)}\,.
\label{U-gen}
\ee
As was shown in Section~3.1, the function $\Phi_{\mybf{q}}(\Mybf{x})$ satisfies
the multi-dimensional Baxter equation \re{Bax-multi}. In this Section we shall
construct the solutions to \re{Bax-multi}. The analysis is based on the relation
between the transition function to the SoV representation, $U_{p,\mybf{x}}(z)$,
and the Baxter ${\mathbb Q}-$operator for the $SL(2,\mathbb{R})$ spin chain.

\begin{figure}[t]
\psfrag{z1}[cc][cc]{$z_1$}
\psfrag{z2}[cc][cc]{$z_2$}
\psfrag{zn1}[cc][cc]{$z_{N-1}$}
\psfrag{zn}[cl][cc]{$z_N$}
\psfrag{wn}[cc][cc]{$\bar w_N$}
\psfrag{w1}[cc][cc]{$\bar w_1$}
\psfrag{w2}[cc][cc]{$\bar w_2$}
\psfrag{y1}[cc][cb]{$y_1$}
\psfrag{y2}[cc][cb]{$y_2$}
\psfrag{www}[cc][cc]{$\scriptstyle\ \ \ \bar w_1=\ldots=\bar w_{N-1}\to\infty$}
\centerline{\epsfxsize16.0cm\epsfbox{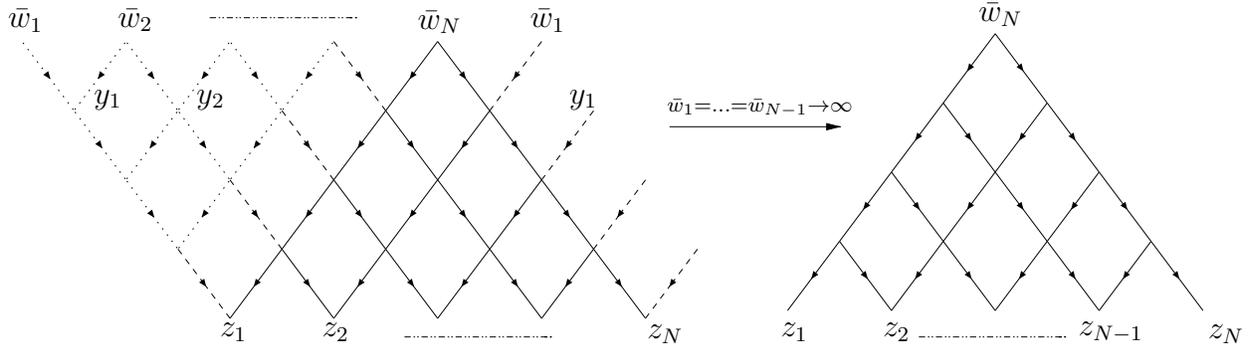}}
\vspace*{0.5cm}
\caption[]{Reduction of the product of $(N-1)$ Baxter $\mathbb{Q}-$operators
to the pyramid diagram. The dotted lines shrink into a point at $\bar w_1=\bar
w_2=\ldots=\bar w_{N-1}$. The dashed lines are reduced to a power of $\bar w_1$
in the limit $w_1\to\infty$. }
\label{reduction}
\end{figure}

By the definition~\cite{Baxter}, the ${\mathbb Q}-$operator acts on the quantum
space of the model $\mathcal{V}_N$, depends on the spectral parameter $u$ and
satisfies the following conditions. It commutes with the transfer matrix
$\widehat t_N(u)$ and with itself for different values of the spectral parameter,
$[\widehat t_N(u),{\mathbb Q}(v)]=[{\mathbb Q}(u),{\mathbb Q}(v)]=0$ and fulfil
the operator Baxter equation
\be\label{Bax-E}
\widehat t_N(u)\,{\mathbb Q}(u)=(u+is)^N{\mathbb Q}(u+i)+(u-is)^N{\mathbb Q}(u-i)\,.
\ee
The eigenstates of the model \re{Sch} diagonalize the ${\mathbb Q}-$operator
\be
{\mathbb Q}(u)\Psi_{p,\mybf{q}}(z_1,\ldots,z_N)=Q_{\mybf{q}}(u)
\Psi_{p,\mybf{q}}(z_1,\ldots,z_N)\,,
\label{Bax-eig}
\ee
and the corresponding eigenvalues $Q_{\mybf{q}}(u)$ satisfy the same equation
\re{Bax-E}. The Schr\"odinger equation \re{Sch} is equivalent to \re{Bax-eig} and
the energy $E_{\mybf{q}}$ can be calculated as a logarithmic derivative of
$Q_{\mybf{q}}(u)$ at $u=\pm is$~\cite{FK,K1,SD}.

The Baxter ${\mathbb Q}-$operator for the homogenous $SL(2,\mathbb{R})$ spin
magnet was constructed as an integral operator in Ref.~\cite{SD}
\be\label{Q-int}
\left [{\mathbb Q}(u)\,\Psi\right](z_1,\ldots,z_N)=\int \mathcal{D}^N w\,
  Q_u(z_1,\ldots,z_N;\bar w_1,\ldots,\bar w_N)\, \Psi(\bar w_1,\ldots,\bar w_N)\,.
\ee
and the kernel was calculated following the Pasquier--Gaudin approach~\cite{PG}
as
\be\label{Q-kernel}
 Q_u(\vec z\,;\vec w)=\e^{i\pi s N}\prod_{k=1}^N\,(z_k-\bar
 w_{k+1})^{-s+iu}(z_k-\bar w_k)^{-s-iu}\,,
\ee
where $w_{N+1}=w_1$. The conjugated operator $({\mathbb Q}(u^*))^\dagger$ is
defined in a similar way and its kernel is given by $\lr{Q_{u^*}(\vec w\,;\vec
z)}^*$. Using \re{Q-kernel} one can verify the following relation
\be
({\mathbb Q}(u^*))^\dagger=\mathbb{P}\,{\mathbb Q}(u)={\mathbb
Q}(u)\,\mathbb{P}\,,
\label{DP}
\ee
where $\mathbb{P}$ is the operator of cyclic permutations of $N$ particles,
$\mathbb{P}^N=\II$,
\be
\mathbb{P}\,\Psi_{\mybf{q},p}(z_1,\ldots,z_{N-1},z_N)=\Psi_{\mybf{q},p}(z_2,\ldots,z_N,z_1)
=\e^{i\theta_{\mybf{q}}}\Psi_{\mybf{q},p}(z_1,\ldots,z_{N-1},z_N)\,.
\label{P}
\ee
Here in the second relation we took into account that the Hamiltonian of the
model, Eq.~\re{H}, is invariant under the cyclic permutations of particles and,
therefore, its eigenstates possess a definite value of the quasimomentum
$\theta_{\mybf{q}}=2\pi k/N$ with $k\in \mathbb{N}$. It follows from \re{DP} that
the operators $({\mathbb Q}(u^*))^\dagger$ and ${\mathbb Q}(u)$ share the common
set of the eigenfunctions and their eigenvalues satisfy the relation
\be
\lr{Q_{\mybf{q}}(u^*)}^*=\e^{i\theta_{\mybf{q}}}Q_{\mybf{q}}(u)\,.
\label{Q-conj}
\ee
At $u=-is$ the kernel of the $\mathbb{Q}-$operator \re{Q-kernel} coincides with
the kernel of the unity operator \re{K} leading to
\be
\mathbb{Q}(-is)= \mathbb{K}\,,
\ee
so that $Q_{\mybf{q}}(-is)=1$.

Notice that Eq.~\re{Q-kernel} looks similar to the definition of the kernel of
the $\Lambda-$operator, Eq.~\re{La}, which enters into the expression for the SoV
transformation \re{U}. Namely, the expression for $Q_u(\vec z\,;\vec w)\e^{-i\pi
s N}$, Eq.~\re{Q-kernel}, coincides with the $Y-$function, Eq.~\re{Bxal}, and it
differs from the $\Lambda-$function, Eq.~\re{La}, by two factors, $(z_1-\bar
w_{1})^{-s-iu}$ and $(z_N-\bar w_{1})^{-s+iu}$. This suggests that there should
exist a relation between the transition function to the SoV representation
$U_{\mybf{x}}(\vec z\,;\bar w_N)$ and the kernel of the ${\mathbb Q}-$operator.
For the $SL(2,\mathbb{C})$ magnet such relation has been found in
Ref.~\cite{DKM}.

As a starting point, one considers the following transformation
$\Psi\to\Phi^\Omega$, the so-called separating map~\cite{KS}
\ba
\Phi^\Omega(x_1,\ldots,x_{N-1}) &=& \vev{\Omega|\,{\mathbb Q}(x_1)\ldots {\mathbb
Q}(x_{N-1})|\Psi}
\nonumber
\\
&&\hspace*{-30mm}=\int \mathcal{D}^N z_1\ldots \int \mathcal{D}^N z_N\,
\lr{\Omega(\vec z_1)}^* Q_{x_1}(\vec z_1;\vec z_2)
\ldots Q_{x_{N-1}}(\vec z_{N-1};\vec z_N) \Psi(\vec z_N)\,,
\label{SoV-for}
\ea
where $\Omega(z_1,\ldots,z_N)$ is an arbitrary function holomorphic on the
upper-half plane. When applied to the eigenfunction of the model,
$\Psi_{p,\mybf{q}}(z_1,\ldots,z_N)$, it separates the variables for arbitrary
$\Omega(\vec z)$
\be
\Phi^\Omega(x_1,\ldots,x_{N-1})= Q_{\mybf{q}}(x_1)\ldots Q_{\mybf{q}}(x_{N-1})
\vev{\Omega|\Psi_{\mybf{q},p}}\,.
\ee
To reconstruct the eigenfunction $\Psi_{\mybf{q},p}(z_1,\ldots,z_N)$ one has to
invert \re{SoV-for} and define the inverse separating map $\Phi^\Omega\to\Psi$.
Notice that for arbitrary $\Omega(\vec z)$ the transformation \re{SoV-for} is not
unitary and, therefore, it does not correspond to the SoV transformation
\re{U-gen}.
Let us demonstrate however that there exists special state $\Omega(\vec z)$, for
which the separating map \re{SoV-for} becomes unitary. In that case, \re{SoV-for}
coincides with the SoV transformation \re{U-gen} and the inverse separating map
is defined by \re{SoV-gen}.

Following our diagrammatical approach, we use \re{Q-kernel} and represent the
kernel of the product of $(N-1)$ Baxter operators, $\left[{\mathbb Q}(x_1)\ldots
{\mathbb Q}(x_{N-1})\right](\vec z\,; \vec w)$ as the Feynman diagram shown in
Figure~\ref{reduction} to the left. Notice that it contains as a subgraph the
pyramid diagram corresponding to the transition function
$U_{p,\mybf{x}}(\vec{z};\bar w_N)$. It is possible to reduce the whole diagram to
its subgraph as follows. One explores a freedom in choosing the $\vec
w-$coordinates to put $\bar w_1=\bar w_2$ in the upper row of the left diagram in
Figure~\ref{reduction}. The two lines connecting the points $\bar w_1$ and $\bar
w_2$ with the vertex $y_1$ merge into a single line with the index $2s$, which
corresponds (up to a numerical factor) to the reproducing kernel $K(y_1,\bar
w_1)=\e^{i\pi s}(y_1-\bar w_{1})^{-2s}$, defined in Eqs.~\re{K-ker} and \re{K}.
Therefore, one can effectively remove the $y_1-$integration and put $y_1=w_1$
instead. At the diagrammatical level this is equivalent to shrinking into a point
the two lines connecting $\bar w_1$ and $\bar w_2$ with $y_1$. In a similar
manner, choosing $\bar w_1=\ldots=\bar w_{N-1}$ in the upper row of the left
diagram in Figure~\ref{reduction}, one creates an avalanches of simplifications
throughout the diagram which removes $2(k-1)$ lines in the $k$th row shown there
by the dotted lines. The resulting diagram still contains $2(N-1)$ additional
lines (the dashed lines in the Figure~\ref{reduction}), which connect the point
$w_1$ with the vertices along the boundary of the pyramid. For $w_1\to
\infty$ their product scales as $\bar w_1^{-2s(N-1)}$. To get rid of the
additional lines one multiplies the whole diagram by $\bar w_1^{2s(N-1)}$ and
sends $w_1\to\infty$.

In this way, we arrive at the following relation
\be
U_{\mybf{x}}(\vec z\,;\bar w_N)=i^{-(N-1)(N+2)s}\lim_{w_1\to\infty}\bar
w_1^{2s(N-1)}
\left[{\mathbb Q}(x_1)\ldots
{\mathbb Q}(x_{N-1})\right](\vec z\,;\bar w_1,\ldots,\bar w_1,\bar w_N)\,.
\label{U=Q}
\ee
It can be rewritten in an operator form by introducing the special state
\be
\Omega_{\bar w_0,\bar w_N}(z_1,\ldots,z_N)=i^{-(N-1)(N+2)s}
\prod_{k=1}^{N-1}\frac{\e^{i\pi s}\bar w_0^{2s}}{(z_k-\bar w_0)^{2s}}
\cdot \frac{\e^{i\pi s}}{(z_N-\bar w_N)^{2s}}\,,
\ee
which depends on two complex parameters $\bar w_0$ and $\bar w_N$. This state is
normalizable with respect to the $SL(2,\mathbb{R})$ scalar product
\be
\vev{\Omega_{\bar w_0',\bar w_N'}|\Omega_{\bar w_0,\bar w_N}}=
\e^{i\pi N s}({\bar w_0\!}^{-1}-{w_0'}^{-1})^{-2s(N-1)}(w_N'-\bar
w_N)^{-2s}
\label{sc-Omega}
\ee
and, therefore, it belongs to the quantum space of the model. Then, one finds
from \re{U=Q} the following relation between the unitary transformation to the
SoV representation and the product of $(N-1)$ Baxter $\mathbb{Q}-$operators
\be
U_{\mybf{x}}(\vec z\,;\bar w_N)=\lim_{\bar w_0\to\infty}\vev{\vec z|\,{\mathbb
Q}(x_1)\ldots {\mathbb Q}(x_{N-1})|\Omega_{\bar w_0,\bar w_N}}\,,
\label{U-op}
\ee
where it is implied that one has to evaluate the matrix element and take the
limit $\bar w_0\to\infty$ afterwards. Substituting this relation into \re{U-gen}
and taking into account \re{DP}, one obtains
\ba
\vev{w_N,\Mybf{x}|\Psi_{\mybf{q},p}}&=&\lim_{\bar w_0\to\infty}\vev{\Omega_{\bar w_0,\bar w_N}|\,{\mathbb
Q}(x_1)\ldots {\mathbb Q}(x_{N-1})\mathbb{P}^{N-1}|\Psi_{\mybf{q},p}}
\nonumber
\\
&=&c_{\mybf{q}}(p)\cdot\e^{ipw_N}Q_{\mybf{q}}(x_1)\ldots Q_{\mybf{q}}(x_{N-1})\,,
\label{vev-fac}
\ea
where $c_{\mybf{q}}(p)=\e^{-i\theta_{\mybf{q}}}\lim_{\bar
w_0\to\infty}\vev{\Omega_{\bar w_0,\bar w_N=0}|\Psi_{\mybf{q},p}}$. The
$p-$dependence of $c_{\mybf{q}}(p)$ follows from \re{U-gen} as 
$c_{\mybf{q}}(p)=c_{\mybf{q}}\cdot p^{-s(N-2)-1/2}/\Gamma(2s)$. Matching the
second relation in \re{vev-fac} into the r.h.s.\ of \re{U-gen} one obtains that
the wave function in the Separated Variables is given by the product of $(N-1)$
eigenvalues of the Baxter operator \re{Bax-eig} evaluated for real $u=x_k$
\be
\Phi_{\mybf{q}}(\Mybf{x})=c_{\mybf{q}}\, Q_{\mybf{q}}(x_1)\ldots
Q_{\mybf{q}}(x_{N-1})\,,
\label{facQ}
\ee
with the factor $c_{\mybf{q}}$ depending on the normalization of the eigenstates
$\Psi_{\mybf{q},p}$.

It is well-known~\cite{SD,FK,K1} that the eigenvalues of the Baxter operator for
the quantum $SL(2,\mathbb{R})$ magnet, $Q_{\mybf{q}}(u)$, are polynomials in $u$
of degree $h\in \mathbb{N}$ defined by the total spin of the system
\be
\vec S^2\,\Psi_{\mybf{q},p}(\vec z)=(h+Ns)(h+Ns-1)\,\Psi_{\mybf{q},p}(\vec z)\,,
\label{h}
\ee
or equivalently $q_2=-(h+Ns)(h+Ns-1)-Ns(s-1)$ in Eq.~\re{q2}. Then, one finds
from \re{facQ} that the wave function $\Phi_{\mybf{q}}(\Mybf{x})$ is given by the
product of $(N-1)$ polynomials in the separated variables. The eigenfunctions of
the model $\Psi_{\mybf{q},p}(\vec z)$ are orthogonal to each other with respect
to the $SL(2,\mathbb{R})$ scalar product \re{scalar}. Going over to the SoV
representation \re{SoV-gen} and using \re{U-ort}, one can write the orthogonality
condition as
\be
\vev{\Psi_{\mybf{q'},p'}|\Psi_{\mybf{q},p}}
=
\vev{\Phi_{\mybf{q'}}|\Phi_{\mybf{q}}}_{_{\rm SoV}}\delta(p-p')
=\delta(p-p')\,\delta_{\mybf{q},\mybf{q'}}\,,
\ee
where we took into account that the spectrum of the integrals of motion for the
$SL(2,\mathbb{R})$ magnet is discrete~\cite{K1,BDKM}. Here the notation was
introduced for the scalar product in the SoV representation
\be
\vev{\Phi_{\mybf{q'}}|\Phi_{\mybf{q}}}_{_{\rm SoV}}=
\int_{\mathbb{R}^{N-1}} d^{N-1} \Mybf{x}\,\mu(\Mybf{x})
\lr{\Phi_{\mybf{q}'}(x_1,\ldots,x_{N-1}}^*
\Phi_{\mybf{q}}(x_1,\ldots,x_{N-1})\,.
\ee
Together with \re{facQ} and \re{Q-conj} this leads to the orthogonality condition
on the space of solutions to the Baxter equation \re{Bax-E} and \re{Bax-eig}
\be
\int_{\mathbb{R}^{N-1}} d^{N-1} \Mybf{x}\,\mu(\Mybf{x})
\prod_{k=1}^{N-1} Q_{\mybf{q}}(x_k)\,Q_{\mybf{q}'}(x_k)
\sim \delta_{\mybf{q},\mybf{q}'}\,.
\ee
At $N=2$ this condition alone allows us to obtain the solution to the Baxter
equation. Given that the integration measure at $N=2$, Eq.~\re{measure_2},
coincides with the weight function for the continuous Hanh orthogonal
polynomials~\cite{Koekoek}, one finds that $Q_{\mybf{q}}(x)$ equals the same
polynomial
\be
Q_h(x)={}_3F_2\lr{{{4s+h-1,-h,s+ix}\atop{2s,2s}}\bigg|\,1}\,,
\label{Hanh}
\ee
with nonnegative integer $h$ defined in \re{h}. This result is in agreement with
the previous calculations~\cite{FK,K1,SD,BDKM}. For higher $N$ the solutions to
the Baxter equation \re{Bax-E} can be expanded over the $N=2$ solutions \re{Hanh}
with the coefficients satisfying the $N-$term recurrence
relations~\cite{K1,BDKM,AB}.

Substituting \re{facQ} into \re{SoV-gen}, one obtains the integral representation
for the eigenfunctions of the model $\Psi_{\mybf{q},p}(\vec z)$. As shown in
Appendix~\ref{B}, the resulting expression for $\Psi_{\mybf{q},p}(\vec z)$
coincides with the well-known highest-weight representation for the eigenstates
in the Algebraic Bethe Ansatz approach~\cite{QISM,ABA}
\be
\Psi_{\mybf{q},p}(\vec z)=B_N(\lambda_1)\ldots B_N(\lambda_h)\, \Omega_p(\vec z)\,.
\label{ABA}
\ee
Here the operator $B_N(u)$ is the off-diagonal element of the monodromy matrix
\re{Tu} and the state $\Omega_p(\vec z)$ is defined as
\be
\Omega_p\,(\vec z)=\frac{p^{2s-1}\e^{i\pi N s}}{\Gamma(2s)}\int_{\Im w>0} \mathcal{D}w\, \e^{ipw}\prod_{k=1}^N
(z_k-\bar w)^{-2s}\,.
\ee
The parameters $\lambda_1,\ldots,\lambda_h$ are simple roots of the eigenvalues
of the Baxter operator, $Q_{\mybf{q}}(u)=\prod_{k=1}^h (u-\lambda_k)$. They
satisfy the Bethe equations
\be
\lr{\frac{\lambda_k+is}{\lambda_k-is}}^N=\prod_{j=1 \atop j \neq k}^h
\frac{\lambda_k-\lambda_j-i}{\lambda_k-\lambda_j+i}\,,
\label{Bethe-eq}
\ee
which follow from the Baxter equation \re{Bax-E} for $Q_{\mybf{q}}(u)$. The fact
that the two different representations for the eigenstates, Eqs.~\re{SoV-gen} and
\re{ABA}, coincide implies that the Algebraic Bethe Ansatz is complete. We recall
that the quantum space of the $SL(2,\mathbb{R})$ magnet is infinite-dimensional
for an arbitrary number of sites $N$ and, as a consequence, the well-known
combinatorial completeness analysis of the Bethe states
(see Ref.~\cite{KL1} and references therein) is not applicable in that case.

\section{Conclusions}
\label{Summary}

In this paper we have studied the spectral problem for the quantum
$SL(2,\mathbb{R})$ spin magnet within the framework of the Quantum Inverse
Scattering Method. This model has previously emerged in high-energy QCD as
describing the spectrum of anomalous dimensions of composite high-twist
operators.

The central point of our analysis was the construction of the representation of
the Separated Variables for the $SL(2,\mathbb{R})$ spin magnet. Following the
Sklyanin's approach, we presented a general method for constructing the unitary
transformation to the SoV representation. It allowed us to obtain the integral
representation for the eigenfunctions of the model and calculate explicitly the
integration measure defining the scalar product in the SoV representation. We
demonstrated that the language of Feynman diagrams becomes extremely useful in
establishing various properties of the model. In particular, we found that the
kernel of the unitary transformation to the SoV representation takes the
factorized form \re{U}. In terms of Feynman diagrams this implies that the kernel
can be described by the pyramid diagram shown in Figure~\ref{Pyramid}. Notice
that the same diagram has been previously encountered in the construction of the
SoV representation for the $SL(2,\mathbb{C})$ spin magnet in Ref.~\cite{DKM,VL}.
The approach described in this paper can be applied to other quantum integrable
models like the periodic Toda chain~\cite{PG,KL} and the DST model~\cite{KSS},
which represent degenerated cases of the $SL(2,\mathbb{R})$ spin chain.



\begin{figure}[t]
\centerline{\epsfxsize10.0cm\epsfbox{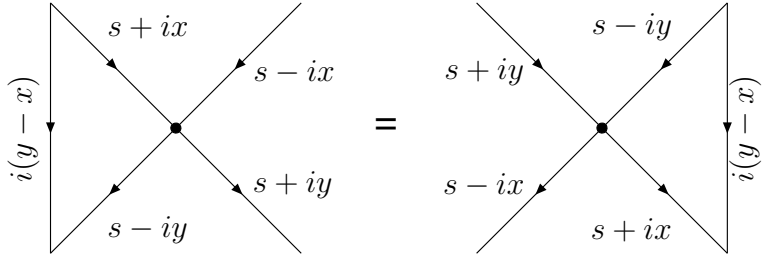}}
\vspace*{0.5cm}
\caption[]{Permutation identity.}
\label{comm-f}
\end{figure}

Remarkably enough, many nontrivial properties of the SoV representation can be
deduced in our approach from a few elementary identities between Feynman diagrams
like the chain and permutation identities shown in Figures~\ref{Chain} and
\ref{comm-f}, respectively. We found that the kernel of the unitary
transformation to the SoV representation is given by the product of the Baxter
$\mathbb{Q}-$operators projected onto a special reference state. This allowed us
to establish the relation between the wave functions in the separated variables
and the eigenvalues of the $\mathbb{Q}-$operator for the $SL(2,\mathbb{R})$ spin
magnet. Using the well-known fact that the latter are polynomials in the spectral
parameter, we demonstrated the equivalence between two different expressions for
the eigenstates obtained within the Algebraic Bethe Ansatz and the SoV method.

So far we have assumed that the spin chain was homogeneous. Another advantage of
our method is that it can be easily extended to the case of the inhomogeneous
closed spin chain. 
For the $SL(2,\mathbb{R})$ chain with the spin in the $k$th site equals $s_k$,
the unitary transformation to the SoV representation is defined by {\it the
same\/} pyramid diagram (see Figure~\ref{Pyramid}) with the $\alpha-$ and
$\beta-$indices modified in the following way. For $s_k\neq s_j$, the indices
take different values for the different lines in the same row. If one denotes by
$\alpha_{nk}$ and $\beta_{nk}$ the indices carried by the $k$th pair of adjacent
lines in the $n$th row from the bottom, ($1\le n\le N-1$ and $1\le k\le N-n$),
then $\alpha_{nk}=s_k-ix_n$ and $\beta_{nk}=s_{n+k}+ix_n$. In addition, in order
to preserve the $SL(2,\mathbb{R})$ invariance of integrals, one has to modify the
integration measure \re{measure} in all vertices of the diagram. Namely, one has
to replace $(2\Im w_{nk})^{2s-2}\to (2\Im w_{nk})^{\alpha_{nk}+\beta_{nk}-2}$ in
the integration measure \re{measure} at the vertex $w_{nk}$, to which the lines
$\alpha_{nk}$ and $\beta_{nk}$ are attached.

\section*{Acknowledgements}

This work was supported in part by the grant 00-01-005-00 of the Russian
Foundation for Fundamental Research (A.M. and S.D.), by the Sofya Kovalevskaya
programme of Alexander von Humboldt Foundation (A.M.) and by the NATO Fellowship
(A.M.).

\begin{figure}[t]
\centerline{\epsfxsize10.0cm\epsfbox{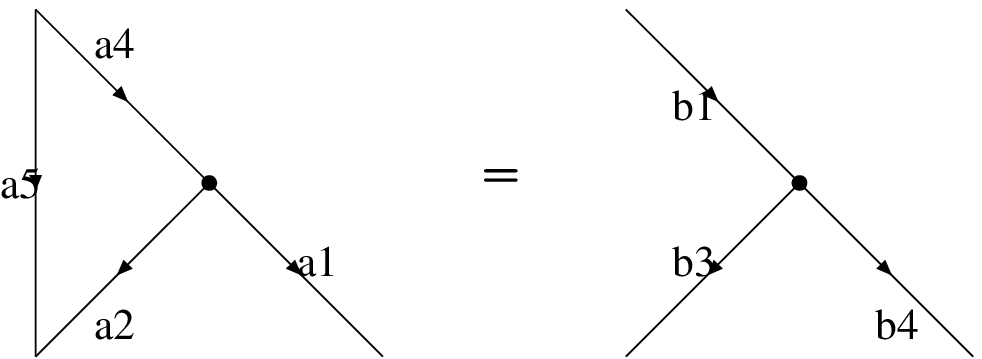}}
\vspace*{0.5cm}
\caption[]{Special case of the permutation identity. It is obtained
{}from Figure~\ref{comm-f} by sending one of the external points to infinity.}
\label{amp1}
\end{figure}
\appendix
\renewcommand{\theequation}{\Alph{section}.\arabic{equation}}
\setcounter{table}{0}
\renewcommand{\thetable}{\Alph{table}}

\section{Appendix: Feynman diagram technique}
\label{A}

In this Appendix we collect some useful formulae, which were used to prove the
permutation identity shown in Figure~\ref{comm-f} and to calculate of the
integration measure \re{mfi}.

\begin{figure}[t]
\psfrag{a}[cc][cc]{$\beta$}
\psfrag{b}[cc][cc]{$\alpha$}
\psfrag{ab}[cc][cc]{$\alpha+\beta-2s$}
\psfrag{text}[cc][cc]{$=\ \ \ \ \ \ \ a(\alpha,\beta)\ \ \ \ \times $}

\centerline{\epsfxsize14.0cm\epsfbox{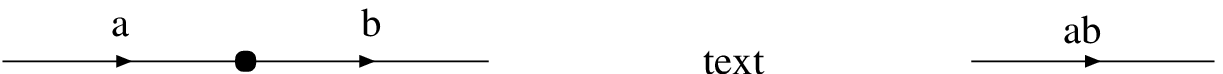}}
\vspace*{0.5cm}
\caption[]{Chain relation.}
\label{Chain}
\end{figure}

To evaluate the $SL(2,\mathbb{R})$ integrals we use the identity
\be
\int_{\Im w >0} \mathcal{D} w  \e^{i p w-ip'\bar w} = \delta(p-p')
\,p^{1-2s}\cdot\Gamma(2s)\,,
\label{delta-f}
\ee
with $\mathcal{D} w$ defined in \re{measure}, $\bar w=w^*$ and $p>0$. The
momentum representation for function $\Psi(z)$ holomorphic in the upper-half
plane is defined as
\be
\Psi(z)=\int_0^\infty dp\,\e^{ipz} \widetilde \Psi(p)\,,\qquad
\widetilde \Psi(p)=\frac{\theta(p)p^{2s-1}}{\Gamma(2s)}\int_{\Im z>0} \mathcal{D}
z\,\e^{-ip\bar z}\Psi(z)\,.
\label{mom-rep}
\ee
Applying the integral representation for the propagators
\be
\frac{1}{(z-\bar w)^\alpha}=\frac{\e^{-i\pi
\alpha/2}}{\Gamma(\alpha)}\,\int_0^{\infty} dp\,\e^{ip\,(z-\bar w)} \,
p^{\alpha-1}\,,
\label{alpha-rep}
\ee
which is valid for $\Im(z-\bar w)>0$, one finds
\ba
&&
\int_{\Im w >0} \mathcal{D} w\, \frac{\e^{i p w}}{(z-\bar{w})^{\alpha}} =
 \theta(p)\,p^{\alpha-2s}\e^{i p z}\cdot\e^{-i\pi\alpha/2}\frac{\Gamma(2s)}{\Gamma(\alpha)}\,,
\label{p-rep}
\\
&&
\int_{\Im w >0} \mathcal{D}w\,{(z-\bar w)^{-\alpha} (w-\bar v)^{-\beta}~=~
{a(\alpha,\beta)}\,(z-\bar v)^{-\alpha-\beta+2s}}\,.
\label{chain-h}
\ea
Here the notation was introduced for
\be\label{As}
a(\alpha,\beta)=\e^{-i\pi s}\,
\frac{\Gamma(\alpha+\beta-2s)\Gamma(2s)}{\Gamma(\alpha)\Gamma(\beta)}\,.
\ee
Eq.\re{chain-h} is the ``chain relation'' shown in Figure~\ref{Chain}. From
Eq.\re{p-rep} and \re{mom-rep} one finds that
\be
\int_{\Im w >0} \mathcal{D}w\,\frac{\e^{i\pi s}}{(z-\bar w)^{2s}}\Psi(w)=
\Psi(z)\,.
\label{K-ker}
\ee
Using the Feynman parameterization
$$
x_1^{-\alpha}x_2^{-\beta}=\frac{\Gamma(\alpha+\beta)}{\Gamma(\alpha)\Gamma(\beta)}
\int_0^1d\tau \tau^{\alpha-1}(1-\tau)^{\beta-1}[\tau
x_1+(1-\tau)x_2]^{-\alpha-\beta}
$$
and applying \re{K-ker}, one finds that for $\alpha+\beta=2s$~\cite{SD}
\be
\int_{\Im w >0} \mathcal{D}w\,\frac{\e^{i\pi s}}{(z_1-\bar w)^{\alpha}
(z_2-\bar
w)^{\beta}}\Psi(w)=\frac{\Gamma(2s)}{\Gamma(\alpha)\Gamma(\beta)}
\int_0^1d\tau \tau^{\alpha-1}(1-\tau)^{\beta-1}\Psi(\tau z_1+(1-\tau)z_2)\,.
\label{drift}
\ee
The permutation relation shown in the Figure~\ref{comm-f} involves the Feynman
integral
\be
I(z,\bar v\,;x_1,x_2)=
\int_{\Im w>0} \mathcal{D}w \frac{1}{(w-\bar v_1)^{\alpha_1}(w-\bar v_2)^{\beta_1}
(z_1-\bar w)^{\beta_2}(z_2-\bar w)^{\alpha_2}}\,,
\label{four}
\ee
with $\alpha_k=s-ix_k$ and $\beta_k=s+ix_k$. One applies \re{drift} for
$\Psi(w)=(w-\bar v_1)^{-\alpha_1}(w-\bar v_2)^{-\beta_1}$ and calculates the
integral in terms of ${}_2F_1-$hypergeometric function as
\be\label{four-2}
I(z,\bar v\,;x_1,x_2)=\e^{-i\pi s}\,(z_2-\bar v_2)^{-\alpha_2} (z_1-\bar
v_1)^{-\alpha_1}\,(z_1-\bar v_2)^{\alpha_1+\alpha_2-2s}\,
{}_2F_1\left(\alpha_1,\alpha_2;2s;\xi\right)\,,
\ee
with $\xi= {(z_1-z_2)(\bar v_1-\bar v_2)}/[(z_1-\bar v_1)(z_2-\bar v_2)]$. {}From
here, one gets for $x_1=x$ and $x_2=y$
\be
\label{comm-i}
(z_2-\bar v_2)^{i(x-y)}I(z,\bar v\,;x,y)=(z_1-\bar v_1)^{i(x-y)} I(z,\bar
v\,;y,x)\,.
\ee
Multiplying the both sides of Eq.~(\ref{comm-i}) by $\bar v_2^{s-ix}$ and taking
the limit $v_2\to\infty$, one obtains the reduced permutation relation shown in
Figure~\ref{amp1}.

Finally, let us verify the completeness conditions \re{completeness} at $N=2$. We
use \re{alpha-rep} to rewrite the transition function $U_{p,x}(z_1,z_2)$ defined
in \re{N=2} and \re{B-ei} in the momentum representation as
\be
U_{p,x}(z_1,z_2)=\frac{\e^{-i\pi
s}\Gamma(2s)p^{1/2}}{\Gamma(s-ix)\Gamma(s+ix)}\int_0^\infty \frac{dp_1}{
 p_1^{1-s}}
\int_0^\infty\frac{dp_2}{p_2^{1-s}}\lr{\frac{p_2}{p_1}}^{ix}\e^{ip_1z_1+ip_2z_2}\,
\delta(p_1+p_2-p)\,.
\label{U-mom}
\ee
At $N=2$ the pyramid diagram in Figure~\ref{Pyramid} consists of two lines. The
variables $p_1$ and $p_2$ have the meaning of the momenta that flow along these
lines. Using similar expression for the conjugated function
$\lr{U_{p,x}(w_1,w_2)}^*$ as a double integral over the momenta $p_1'$ and
$p_2'$, one examines the product $\mu(x)U_{p,x}(z_1,z_2)\lr{U_{p,x}(w_1,w_2)}^*$
with the integration measure $\mu(x)$ given by \re{measure_2}. One notices that
the $x-$dependence of the integration measure is compensated by the
$\Gamma-$functions entering the r.h.s.\ of \re{U-mom}. As a result, the $p-$ and
$x-$integration in \re{completeness} can be easily performed leading to the
following combination of the $\delta-$functions
\be
\delta(p_1+p_2-p_1'-p_2')\delta\lr{\ln\frac{p_1p_2'}{p_2p_1'}}=
\frac{p_1p_2}{p_1+p_2}\delta(p_1'-p_1)\delta(p_2'-p_2)\,.
\ee
Combining together all factors one evaluates the l.h.s.\ of \re{completeness} as
\be
\int_0^\infty dp_1\int_0^\infty dp_2\, \frac{(p_1p_2)^{2s-1}}{\Gamma^2(2s)}\e^{ip_1(z_1-\bar
w_1)+ip_2(z_2-\bar w_2)}=\frac{\e^{i\pi s}}{(z_1-\bar w_1)^{2s}}\frac{\e^{i\pi
s}}{(z_2-\bar w_2)^{2s}}\,.
\ee
This expression coincides with the r.h.s.\ of \re{B-ei} at $N=2$.

\section{Appendix: Relation to the Algebraic Bethe Ansatz}
\label{B}

In this Appendix we demonstrate an equivalence between two different
representations for the eigenstates of the $SL(2,\mathbb{R})$ magnet,
Eqs.~\re{SoV-gen} and \re{ABA}, obtained within the SoV method and the Algebraic
Bethe Ansatz, respectively.

To begin with, one constructs the state equal to the product of the highest
weights in $N$ sites
\be
\Omega(\vec z)=\prod_{k=1}^N  {\e^{i\pi s}}{z_k^{-2s}}\,,\qquad S_k^+ \,\Omega(\vec z)=0\,.
\label{pseudo}
\ee
It is analogous to a pseudovacuum state for compact spin magnets. This state
annihilates the lower off-diagonal matrix element of the Lax operator \re{Lax}
and, therefore, diagonalizes the transfer matrix \re{tu}
\be
\widehat t(u) \Omega(\vec z)=\left[(u+is)^N+(u-is)^N\right]\Omega(\vec z)\,.
\label{t-pseudo}
\ee
Due to the $SL(2)$ invariance of the transfer matrix, $[\widehat t(u),\vec S]=0$
with $\vec S=\sum_{k=1}^N \vec S_k$ being the total spin of the magnet, the state
$\e^{\bar w S^-}\Omega(\vec z)=\Omega(z_1-\bar w,\ldots,z_N-\bar w)$ also
verifies \re{t-pseudo}%
\footnote{Notice that in distinction with \re{pseudo}, the state
$\Omega(z_1-\bar w,\ldots,z_N-\bar w)$ has a finite $SL(2,\mathbb{R})$ norm for
$\Im\bar w<0$.}. Applying \re{mom-rep}, one performs its Fourier transformation
\be
\Omega_p(\vec z)=\frac{p^{2s-1}}{\Gamma(2s)}\int_{\Im w>0} \mathcal{D} w\, \e^{ipw}
\Omega(z_1-\bar w,\ldots,z_N-\bar w)\,.
\ee
The state $\Omega_p(\vec z)$ carries the total momentum $p>0$, diagonalizes the
transfer matrix \re{t-pseudo} and satisfies the following relations
\be
iS^- \,\Omega_p(\vec z)=p\,\Omega_p(\vec z)\,,\qquad
\vec S^2\,\Omega_p(\vec z)=s(s-1)\Omega_p(\vec z)\,.
\label{pseu}
\ee
Its $SL(2,\mathbb{R})$ norm can be easily calculated from \re{p-rep} and
\re{delta-f} as
\be
\vev{\Omega_{p'}|\Omega_p}=\int \mathcal{D}^N z\, \lr{\Omega_{p'}(\vec z)}^*\Omega_{p}(\vec z)
=\delta(p-p')\frac{p^{2Ns-1}}{\Gamma(2Ns)}\,.
\label{sc-pseu}
\ee
This implies that the state $\Omega_p(\vec z)$ belongs to the quantum space of
the model. Then, it follows from Eqs.~\re{t-pseudo} and \re{pseu} that, in virtue
of complete integrability of the model, $\Omega_p(\vec z)$ is an eigenstate of
the Hamiltonian \re{H}. Its total $SL(2,\mathbb{R})$ spin can be found from
\re{pseu} and \re{h} to be $h=0$. According to \re{J} and \re{H}, the energy of
this state is $E_{\mybf{q}}=0$ since $(J_{n,n+1}-2s)\Omega_p(\vec z)=0$ for
$n=1,\ldots,N$. The values of the corresponding integrals of motion $\Mybf{q}$
can be found by matching \re{t-pseudo} into \re{tu-exp}.

Let us transform the state $\Omega_p(\vec z)$ into the SoV representation. One
finds from \re{vev-fac}
\be
\vev{w_N,\Mybf{x}|\Omega_p}=\lim_{w_0\to\infty} \vev{\Omega_{\bar w_0,\bar w_N}
|\,\mathbb{Q}(x_1)\ldots \mathbb{Q}(x_{N-1})|\Omega_p}\,,
\label{vev-Omega}
\ee
since $\mathbb{P}\,\Omega_p(\vec z)=\Omega_p(\vec z)$ (see \re{P} and
\re{pseudo}). Notice that $\Omega(z_1-\bar w,\ldots,z_N-\bar w)$ is proportional
to the reproducing kernel \re{K} and its convolution with the kernel of the
$\mathbb{Q}-$operator, Eqs.~\re{Q-int} and \re{Q-kernel}, can be easily performed
leading to $\mathbb{Q}(u)\ket{\Omega_p}=\ket{\Omega_p}$. In this way, one finds
{}from \re{vev-Omega} and \re{sc-Omega}
\be
\vev{w_N,\Mybf{x}|\Omega_p}=
\e^{ip
w_N}\frac{p^{2s-1}}{\Gamma(2s)} \e^{i \pi s (N-1)(N+4)/2} \,.
\ee
Together with the completeness condition \re{completeness} this leads to
\be
\Omega_p(\vec z)=p^{Ns-1/2}i^{(N-1)(N+4)s} \int d^{N-1}\Mybf{x}\,\mu(\Mybf{x})
U_{p,\mybf{x}}(\vec z)\,.
\label{Omega-SoV}
\ee
Substituting this relation into \re{sc-pseu} and taking into account \re{U-ort},
one gets
\be
\int_{\mathbb{R}^{N-1}} d^{N-1}\Mybf{x}\,\mu(\Mybf{x})=\frac1{\Gamma(2Ns)}\,,
\ee
where the measure $\mu(\Mybf{x})$ is given by \re{mfi}.

Let us apply the operator $B_N(\lambda_1)\ldots B_N(\lambda_h)$ to the both sides
of \re{Omega-SoV}, with $B_N$ being the off-diagonal matrix element of the
monodromy operator \re{Tu} and $\lambda_k$ some parameters. By the definition
\re{B-pol}, the transition function $U_{p,\mybf{x}}(\vec z)$ diagonalizes the
operator $B_N$
\be
B_N(\lambda_1)\ldots B_N(\lambda_h)U_{p,\mybf{x}}(\vec z)= (-1)^{h(N-1)} p^h
\prod_{j=1}^{N-1}\left\{\prod_{k=1}^h (x_j-\lambda_k)\right\}U_{p,\mybf{x}}(\vec z)
\,.
\label{B-stack}
\ee
Let us choose $\lambda_1,\ldots,\lambda_h$ to be roots of the eigenvalue of the
Baxter $\mathbb{Q}-$operator, $Q_{\mybf{q}}(\lambda_k)=0$, so that the expression
in the parenthesis coincides with $Q_{\mybf{q}}(x_j)$. Finally, one gets from
\re{Omega-SoV} and \re{B-stack}
\be
B_N(\lambda_1)\ldots B_N(\lambda_h)\Omega_p(\vec z)=c(p)\int
d^{N-1}\Mybf{x}\,\mu(\Mybf{x}) U_{p,\mybf{x}}(\vec
z)\prod_{j=1}^{N-1}Q_{\mybf{q}}(x_j)
\label{final}
\ee
with the normalization factor $c(p)=p^{Ns+h-1/2}(-1)^{h(N-1)}i^{(N-1)(N+4)s}$.
The l.h.s.\ of \re{final} coincides with the well-known expression for the
eigenstates in the Algebraic Bethe Ansatz, whereas the r.h.s.\ defines the
integral representation for the same eigenstate in the SoV representation,
Eq.~\re{SoV-gen}.

\end{document}